\documentclass{aastex631}

\usepackage{float}
\usepackage{chngpage}
\usepackage{amsmath}
\usepackage{tikz}
\usetikzlibrary{arrows,calc,positioning}
\usepackage[normalem]{ulem}
\usepackage{soul}
\usepackage{xcolor}

\tikzstyle{intt}=[draw,text centered,minimum size=6em,text width=5.25cm,text height=0.34cm]
\tikzstyle{intl}=[draw,text centered,minimum size=2em,text width=2.75cm,text height=0.34cm]
\tikzstyle{int}=[draw,minimum size=2.5em,text centered,text width=3.5cm]
\tikzstyle{intg}=[draw,minimum size=3em,text centered,text width=6.cm]
\tikzstyle{sum}=[draw,shape=circle,inner sep=2pt,text centered,node distance=3.5cm]
\tikzstyle{summ}=[drawshape=circle,inner sep=4pt,text centered,node distance=3.cm]
\usetikzlibrary{shapes.geometric, arrows}
\tikzstyle{arrow} = [thick,->,>=stealth]

\newcommand{\NSearched}{388}
\newcommand{\NObserved}{902}
\newcommand{\NSpectraReduced}{6461}
\newcommand{\NSpectraSearched}{1983}
\newcommand{\DataVolume}{9.545} 
\newcommand{\NTess}{51}

\newcommand{\LemkW}{84}
\newcommand{\medianDistance}{78.5}
\newcommand{\vsini}{Vsin$(i)$}

\newcommand\soutpars[1]{\let\helpcmd\sout\parhelp#1\par\relax\relax}
\long\def\parhelp#1\par#2\relax{%
  \helpcmd{#1}\ifx\relax#2\else\par\parhelp#2\relax\fi%
}

\shorttitle{APF Laser Search}
\shortauthors{Zuckerman et al.}
\graphicspath{{./}{images/}}

\begin{document}

\title{The Breakthrough Listen Search for Intelligent Life: A Laser Search Pipeline for the Automated Planet Finder}


\newcommand{\FOOTBALLERS}{Department of Physics and Astronomy, University of Manchester, UK}
\newcommand{\KZA}{University of Malta, Institute of Space Sciences and Astronomy, Msida, MSD2080, Malta}

\correspondingauthor{Zuckerman}
\email{anna\_zuckerman@alumni.brown.edu}

\newcommand{\UCB}{Department of Astronomy,  University of California Berkeley, Berkeley CA 94720}
\newcommand{\seti}{SETI Institute, Mountain View, California}
\newcommand{\NAU}{Department of Physics and Astronomy, Northern Arizona University, 527 S Beaver St, Flagstaff, AZ 86011}
\newcommand{\Brown}{Department of Physics, Brown University, Box 1843, 182 Hope St., Providence, RI 02912}
\newcommand{\ICRAR}{International Centre for Radio Astronomy Research, 1 Turner Ave, Bentley, WA 6102, Australia}

\author[0000-0002-2412-517X]{Anna Zuckerman}
\affiliation{\UCB}
\affiliation{\Brown}

\author[0000-0002-4962-2543]{Zoe Ko}
\affiliation{\UCB}

\author[0000-0002-0531-1073]{Howard Isaacson}
\affiliation{\UCB}
\affiliation{Centre for Astrophysics, University of Southern Queensland, Toowoomba, QLD, Australia}

\author[0000-0003-4823-129X]{Steve Croft}
\affiliation{\UCB}
\affiliation{\seti}

\author[0000-0003-2783-1608]{Danny Price}
\affiliation{\UCB}
\affiliation{\ICRAR}

\author[0000-0002-7042-7566]{Matt Lebofsky}
\affiliation{\UCB}

\author[0000-0003-2828-7720]{Andrew Siemion}
\affiliation{\UCB}
\affiliation{\seti}
\affiliation{\FOOTBALLERS}
\affiliation{\KZA}

\begin{abstract}
The Search for Extraterrestrial Intelligence (SETI) has traditionally been conducted at radio wavelengths, but optical searches are well-motivated and increasingly feasible due to the growing availability of high-resolution spectroscopy. We present a data analysis pipeline to search Automated Planet Finder (APF) spectroscopic observations from the Levy Spectrometer for intense, persistent, narrow bandwidth optical lasers. We describe the processing of the spectra, the laser search algorithm, and the results of our laser search on \NSpectraSearched~spectra of \NSearched\ stars as part of the Breakthrough Listen search for technosignatures. We utilize an empirical spectra-matching algorithm called \texttt{SpecMatch-Emp} to produce residuals between each target spectrum and a set of best-matching catalog spectra, which provides the basis for a more sensitive search than previously possible. We verify that \texttt{SpecMatch-Emp} performs well on APF-Levy spectra by calibrating the stellar properties derived by the algorithm against the \texttt{SpecMatch-Emp} library and against Gaia catalog values. We leverage our unique observing strategy, which produces multiple spectra of each target per night of observing, to increase our detection sensitivity by programmatically rejecting events which do not persist between observations. With our laser search algorithm we achieve a sensitivity equivalent to the ability to detect an \LemkW\,kW laser at the median distance of a star in our dataset (\medianDistance\,ly). We present the methodology and vetting of our laser search, finding no convincing candidates consistent with potential laser emission in our target sample.

\end{abstract}


\section{Introduction} \label{sec:intro}

The Search for Extraterrestrial Intelligence (SETI) is, in the most general sense, a search for features in astronomical data that cannot be easily explained through known astrophysics, and which therefore might carry information about the existence of advanced life beyond Earth. 
The Breakthrough Listen (BL) Initiative, launched in 2015, is the most extensive, comprehensive, and sensitive SETI search to date \citep{isaacson2017breakthrough, Worden2017}. BL is an international and interdisciplinary collaboration of astronomers working to answer fundamental questions about the nature and uniqueness of life in the Universe using both radio and optical telescopes at observatories across the world. In addition to producing a comprehensive database of radio spectrograms and optical spectra (available publicly through the Open Data Archive; \citealt{Lebofsky2019}), the BL search has resulted in important auxiliary studies, for instance of fast radio bursts \citep{Price2019,Gajjar2018} and dark matter \citep{2021JPhCS2156a2026K,2022arXiv220208274F}.

Despite the fact that optical lasers offer several clear advantages over radio signals from the perspectives both of information transmission and of detectability \citep{Lipman2019}, SETI efforts have historically been focused on radio wavelengths \citep{Sheikh2021,Siemion2013,Tarter2011}. First proposed by \citet{SchwartzTownes1961}, optical lasers have the potential to provide a powerful form of interstellar communication capable of outshining a host star in a narrow wavelength range \citep{TellisMarcy2017}. Lasers are energy efficient and present a relatively low risk of eavesdropping \citep{TellisMarcy2017} due to their narrow opening angle relative to radio frequencies. Extinction in the ISM is lower at optical (compared with shorter) wavelengths \citep{Rather1991} and lower background is present at optical (compared with radio) wavelengths both from Earth and cosmic interference sources \citep{RossMonteKingsley2011}. It has been shown that, even under the limitations of current human technology, transmission and detection of lasers over interstellar distances would be possible \citep{TellisMarcy2017,ClarkCahoy2018,Lipman2019, GertzMarcy2022}. It is reasonable to hypothesize that intelligent civilizations, if present, might either intentionally direct lasers towards Earth as a means of communication, or else inadvertently signal their presence while using lasers for other purposes (e.g., telescope guiding, communication between spacecraft, or laser propulsion). This laser emission is then theoretically detectable by human telescopes. The strongest astrophysical source of narrow emission in the optical range is the 8446\,\AA~oxygen line, which is outside our searched wavelength region. Earth-bound sources of laser-like emission features such as the reflected light from sodium street lights are detected and vetted in our search for technosignatures.

We present a new method for identifying laser emission in stellar spectra, and use it to search APF-Levy high-resolution spectra. Previous laser searches have analyzed stellar spectra in either 2D echellogram  form \citep{TellisMarcy2015} or in reduced, 1D form \citep{TellisMarcy2017}. We conduct our laser line search by matching each target APF spectrum to the \texttt{SpecMatch-Emp} catalog of known spectra, and subtracting away the linear combination of best-matching spectra to produce a residual. We use the \texttt{SpecMatch-Emp} model \citep{Yee2017} to perform this matching process. The resulting residual is then free of any features present in both the target spectrum and matched spectra, such as stellar absorption (and less commonly, emission) features, which can otherwise obscure a laser emission line. Searching the residuals provides a modest increase in sensitivity compared with searching the target spectra directly in regions near stellar absorption features, by revealing potential emission lines which fall into those features. We perform a laser line search both on the target spectra themselves, and on these residuals, in order to have a standard against which to compare the residual search. Adoption of a SETI-focused observing plan which collects observations in sets of three allows us to achieve a higher sensitivity by efficiently rejecting the most common types of false positives. Past works, such as \cite{TellisMarcy2017} (though not \citealt{TellisMarcy2015}), reject false positives by requiring candidate events fall on the stellar trace. Effective rejection of false positives using consecutive observations permits us to omit this requirement, and retain sensitivity to the full extracted spatial range of each order.

Section~\ref{sec:data} describes the targets and data format, and provides a brief overview of the pre-processing steps. Section~\ref{sec:methods} presents the algorithms used to process APF spectra into the final format used for the laser search and to search the processed spectra for laser emission, and describes the results from a calibration of the algorithm using derived stellar properties. In section~\ref{sec:results}, we identify and classify the most promising candidates, and discuss our detection sensitivity limits. Finally, section~\ref{sec:conclusion} provides a comparison of our laser search to previously published results and makes suggestions for future work.

\section{Data} \label{sec:data}
\subsection{Observation Program and Target Selection} \label{subsec:targets}
The Automated Planet Finder covers wavelengths between 3740 and 9700\,\AA~and has an average resolution of 95000 \citep{radovan2014automated}. In section~\ref{subsec:preprocessing}, we discuss our limitation to the 4997.10–5899.99\,\AA~ wavelength region covered by the spectral library. Observations for Breakthrough Listen started in 2016 and observing time is pooled with the California Planet Search (CPS) using a queue system managed by UC Santa Cruz. Initially observing with 10\% of telescope time, BL now utilizes 20 hours per semester allocated to BL targets as of 2022 January. Spectra are taken with a decker size of $1 \arcsec \times 5 \arcsec$. 

We collected \NSpectraReduced\ APF spectra of \NObserved\ targets totaling \DataVolume\,GB of data. Observations are typically collected in sets of three consecutive exposures, although the reality of observing does not always allow this. Exposure times range from a few minutes to twenty minutes per spectrum, with a median of 12.1 minutes. The resulting set of three exposures per star results in roughly 25 minutes of continuous observation with one minute of readout between observations. Use of the exposure meter limits the SNR of each spectrum to 200 per pixel or the maximum exposure time, whichever is achieved first. 
Our laser search analysis is focused on observations of \NSearched\ stars that were observed in sets of three spectra which are amenable to \texttt{SpecMatch-Emp} analysis (see section \ref{subsec:SM-emp}).

Our sample consists of a subset of the Breakthrough Listen sample \citep{isaacson2017breakthrough} and a selection of TESS identified planet candidates, and after imposing data cuts described below, we ultimately search \NSearched\ stars. We focus our observations on stars that were not previously analyzed in \cite{TellisMarcy2017}, namely  those that are not amenable to precise radial velocities and are not Kepler identified planet candidates. Our sample of stars, with a limiting distance of 50 pc, is supplemented with a set of \NTess\ Transiting Exoplanet Survey Satellite (TESS) identified transiting planet candidates. Our set of 128 TESS targets are brighter than V = 10, have planet radii between planet radii $0.5 < R_p < 4.0$ and orbital periods between 0.5--100 days. We choose TESS targets above declination of -20, and while the APF can guide on stars fainter than V=10, we prioritize brighter targets to achieve better SNR in each spectrum. We augment our randomly oriented target set with a set of transiting targets for a few reasons. The simplest is that these targets are known to host confirmed exoplanets. In addition, there have been several past SETI works focusing on transiting planets in the radio (for instance \citealt{Siemion_2013} and \citealt{Traas_2021}), and we sought to complement these existing radio observations in the optical. Observing transiting systems provide an advantage for detecting intentional beacons because an extraterrestrial intelligence (ETI) may realize it is more detectable to other ETI who can observe its transit, and thus preferentially send beacon signals along the ecliptic plane. An ETI might also realize that observers will aim to study transiting systems and thus be more likely to serendipitously observe a beacon sent along the ecliptic plane.

Past publications have primarily analyzed surveys focused on slowly rotating main-sequence stars of spectral types F, G, K and M using Keck/HIRES observations. Our sample, drawn from \cite{isaacson2017breakthrough}, purposely includes stars outside of the typical targets observed by exoplanet surveys. These include hot stars, stars with nearby companions, highly active, rapidly rotating, and young stars, though the hottest and fastest rotating stars are not amenable to our search algorithm as discussed in section \ref{subsubsec:highvsini}.

From this APF dataset of \NObserved\ targets, we run our laser search routine on stars that are amenable to the creation of residuals. Known spectroscopic binary systems and non-stellar objects such as galaxies are excluded, along with any spectra that were taken with the APF iodine cell in place. The first of these two requirements excludes 618 spectra (96 targets) from the target set and the second excludes an additional 1400 spectra (but only 23 targets, because most stars were observed both with the iodine cell in and out of the light path in separate observations). In order to ensure that the \texttt{SpecMatch-Emp} algorithm performs well on all targets, we also exclude stars with temperature or radius values reported in the Gaia catalog to be outside the range of the \texttt{SpecMatch-Emp} library, or with no temperature or radius reported in the Gaia catalog. This excludes an additional 1609 spectra (291 stars) from the target set. We exclude spectra that are not matched well in the \texttt{SpecMatch-Emp} matching process described in section \ref{subsec:SM-emp}, by requiring that the chi-squared value between the target and combination of best-matching spectra is less than 15. This excludes an additional 721 spectra (68 stars). Finally, in order to require that events are persistent, we require that all spectra from the same night of observing pass the above tests and that each star has at least three observations taken on a single night of observing. This excludes 360 spectra (45 targets). These data cuts leave \NSearched\ stellar targets, corresponding to \NSpectraSearched\ total spectra. Except for the final one, these data cuts represent a limitation of our spectra-matching algorithm, rather than a property of our target sample or observing strategy.

Table~\ref{tab:log_file_snippet} lists a subset of the targets analyzed, including information about the observations used and SNR of each observation. The table includes all observed targets, including some (such as spectroscopic binaries) which are later excluded from the analysis. All spectra, both raw and reduced, are publicly available as FITS files on the Breakthrough Listen Initiative Open Data Archive\footnote{\url{http://seti.berkeley.edu/opendata}}.

\begin{table}[H]
\centering
\begin{tabular}{lllr}
\toprule
Name &          Date observed (UT) &                    Type & SNR \\
HIP21589 & 2020-12-02T08:01:05.12 & High Proper-motion Star & 209 \\
HIP21673 & 2020-12-02T09:19:11.25 &    Spectroscopic Binary & 194 \\
HIP21673 & 2020-12-02T09:22:27.84 &    Spectroscopic Binary & 205 \\
HIP21673 & 2020-12-02T09:25:38.26 &    Spectroscopic Binary & 205 \\
HIP21683 & 2020-12-02T09:27:45.86 & High Proper-motion Star & 109 \\
HIP21683 & 2020-12-02T09:28:55.13 & High Proper-motion Star & 100 \\
HIP21683 & 2020-12-02T09:30:04.43 & High Proper-motion Star & 100 \\
HIP22044 & 2020-12-02T08:04:30.72 & High Proper-motion Star & 152 \\
HIP22044 & 2020-12-02T08:06:39.40 & High Proper-motion Star & 131 \\
HIP22044 & 2020-12-02T08:09:29.42 & High Proper-motion Star & 169 \\
HIP22287 & 2020-11-29T08:01:23.36 & Double or Multiple Star & 112 \\
HIP22287 & 2020-11-29T08:02:57.56 & Double or Multiple Star & 101 \\
HIP22287 & 2020-11-29T08:04:31.42 & Double or Multiple Star & 103 \\
HIP22361 & 2020-11-25T09:50:44.63 & High Proper-motion Star & 118 \\
HIP22361 & 2020-11-25T09:54:14.67 & High Proper-motion Star & 116 \\
HIP22361 & 2020-11-25T09:57:09.08 & High Proper-motion Star &  96 \\
HIP22498 & 2020-11-25T11:02:30.89 &        Eclipsing Binary &  28 \\
HIP22498 & 2020-11-29T08:26:26.48 &        Eclipsing Binary &  68 \\
HIP22498 & 2020-11-29T08:47:10.64 &        Eclipsing Binary &  70 \\
HIP22498 & 2020-11-29T09:07:54.79 &        Eclipsing Binary &  71 \\
HIP22715 & 2020-11-25T09:03:28.18 & High Proper-motion Star &  85 \\
HIP22715 & 2020-11-25T09:24:12.40 & High Proper-motion Star &  87 \\
HIP22715 & 2020-11-25T09:44:56.61 & High Proper-motion Star &  80 \\
HIP22845 & 2020-12-02T11:31:06.97 & High Proper-motion Star & 187 \\
HIP22845 & 2020-12-02T11:33:16.31 & High Proper-motion Star & 184 \\
HIP22845 & 2020-12-02T11:35:30.31 & High Proper-motion Star & 183 \\
HIP23147 & 2020-11-29T09:54:06.46 & High Proper-motion Star &  29 \\
HIP23147 & 2020-11-29T10:34:35.80 & High Proper-motion Star &  71 \\
HIP23147 & 2020-11-29T10:55:19.92 & High Proper-motion Star &  70 \\
HIP23147 & 2020-11-29T11:16:04.05 & High Proper-motion Star &  72 \\
HIP23179 & 2020-11-29T10:09:06.62 & High Proper-motion Star & 114 \\
HIP23179 & 2020-11-29T10:10:36.43 & High Proper-motion Star & 111 \\
HIP23179 & 2020-11-29T10:12:05.79 & High Proper-motion Star & 115 \\
HIP25486 & 2020-10-08T12:38:50.37 &           Variable Star &  20 \\
HIP25486 & 2020-10-08T12:59:34.52 &           Variable Star &  36 \\
HIP25486 & 2020-10-08T13:20:18.66 &           Variable Star &  40 \\
HIP29277 & 2020-08-15T07:27:33.34 & High Proper-motion Star &  38 \\
HIP29277 & 2020-08-15T07:48:17.51 & High Proper-motion Star &  37 \\
HIP29277 & 2020-08-15T08:09:01.64 & High Proper-motion Star &  36 \\
HIP29277 & 2020-10-09T11:57:19.98 & High Proper-motion Star &  37 \\
HIP29277 & 2020-10-09T12:18:04.11 & High Proper-motion Star &  39 \\
HIP29277 & 2020-10-09T12:38:48.26 & High Proper-motion Star &  42 \\
HIP29761 & 2020-08-16T08:11:09.74 & High Proper-motion Star &  94 \\
HIP29761 & 2020-08-16T08:31:53.89 & High Proper-motion Star &  92 \\
HIP29761 & 2020-08-16T08:52:38.03 & High Proper-motion Star & 136 \\
HIP64924 & 2020-07-08T04:31:09.36 & High Proper-motion Star &  28 \\
HIP64924 & 2020-07-08T04:31:55.78 & High Proper-motion Star &  25 \\
HIP64924 & 2020-07-08T04:33:10.11 & High Proper-motion Star &  73 \\
HIP65011 & 2020-07-08T04:55:48.68 &    Spectroscopic Binary &  75 \\
HIP65011 & 2020-07-08T05:16:32.52 &    Spectroscopic Binary &  74 \\
\end{tabular}
\caption{Sample columns listing the names and certain properties of each APF observation. The SNR is calculated per-pixel at a wavelength of 5906\,\AA. The full table includes all target stars, as well additional information about each observation.} 
\label{tab:log_file_snippet}
\end{table}

\subsection{Pre-processing} \label{subsec:preprocessing}
\begin{figure*}[ht]
\centering
\includegraphics[width=0.5\textwidth]{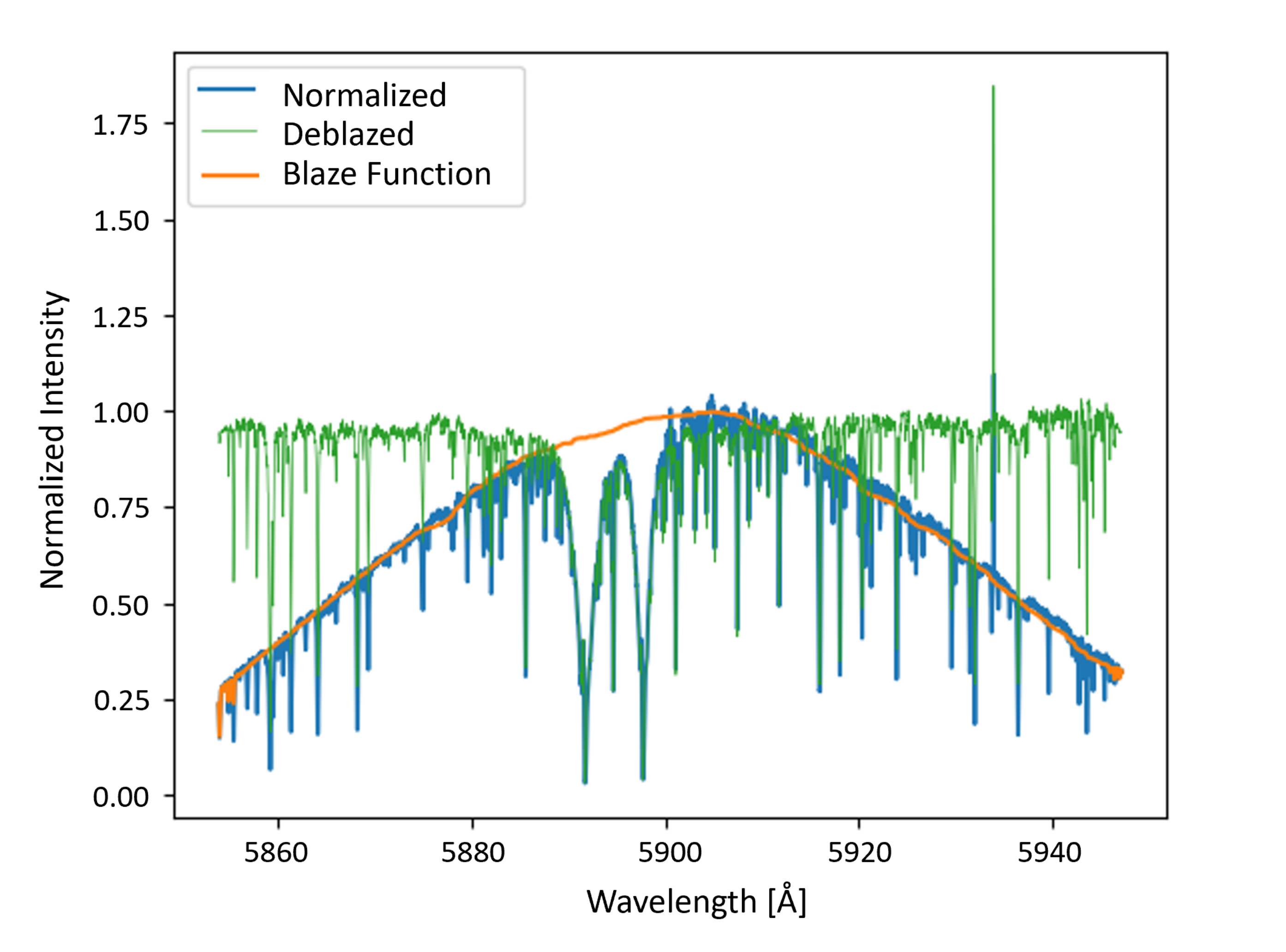}
\caption{An example spectral order (HIP13642, order 45) before and after removal of the blaze function. Deblazing removes the variation of intensity across an order without removing real spectral features. The single bright pixel on right hand side of the spectrum is likely a cosmic ray hit, and is classified during subsequent vetting by the laser search algorithm.}
\label{fig:deblaze}
\end{figure*}

Each target spectrum must be reduced from a raw 2D echellogram to a 1D spectrum, deblazed, continuum-normalized, and combined into a single spectral order to allow for meaningful comparison between observations and stars. First, each raw APF echellogram is reduced to produce a one-dimensional spectrum storing flux values as a function of wavelength in 79 spectral orders. The process is algorithmically very similar to that of \cite{TellisMarcy2017}.
Each spectrum is then normalized in order to compare targets which have different continuum flux values. Individual orders must then be deblazed to remove the dependence of flux as a function of pixel across an order, which is a property of cross-dispersed echelle spectra. We derive blaze functions by smoothing the stacked spectra of rapidly rotating B-type stars, which have relatively few absorption features, and divide these blaze functions from each spectral order (Figure~\ref{fig:deblaze}). Deblazing is imperfect and occasionally results in artifacts, especially near the edges of spectral orders. Some artifacts near order edges can mimic laser emission, but these are vetted by leveraging the requirement that events persist across all observations of the same target.

\noindent Each target spectrum is then resampled onto the wavelength scale of the \texttt{SpecMatch-Emp} library.  The library spectra are limited in resolution by the resolution of the HIRES spectrometer on the Keck telescope, which is about 60000 compared to 90000 for the Levy spectrometer. We interpolate each order in the original spectrum onto a reference wavelength scale. This wavelength scale is defined such that a shift by one pixel represents a constant velocity shift across the spectrum (i.e., ${\rm log}(\frac{\Delta \lambda}{\lambda})$ is constant across the spectrum). The resampled spectrum is within visible region from 4997.10--5899.99\,\AA, and the resolution varies from 0.017\,\AA\ per pixel to 0.021\,\AA\ per pixel across the spectrum. This decrease in wavelength coverage from the full range of the APF is due to the limiting wavelength range of the \texttt{SpecMatch-Emp} library. We flatten the 22 orders of interest into a single 1-dimensional spectrum by averaging the values of pixels which overlap between two orders. The combined, deblazed, normalized and resampled spectra are then ready to be either searched directly for laser emission, or input into \texttt{SpecMatch-Emp}. 

\section{Methods} \label{sec:methods}

\subsection{\texttt{SpecMatch-Emp}} \label{subsec:SM-emp}

\begin{figure*}[ht]
\centering
\includegraphics[width=0.83\textwidth]{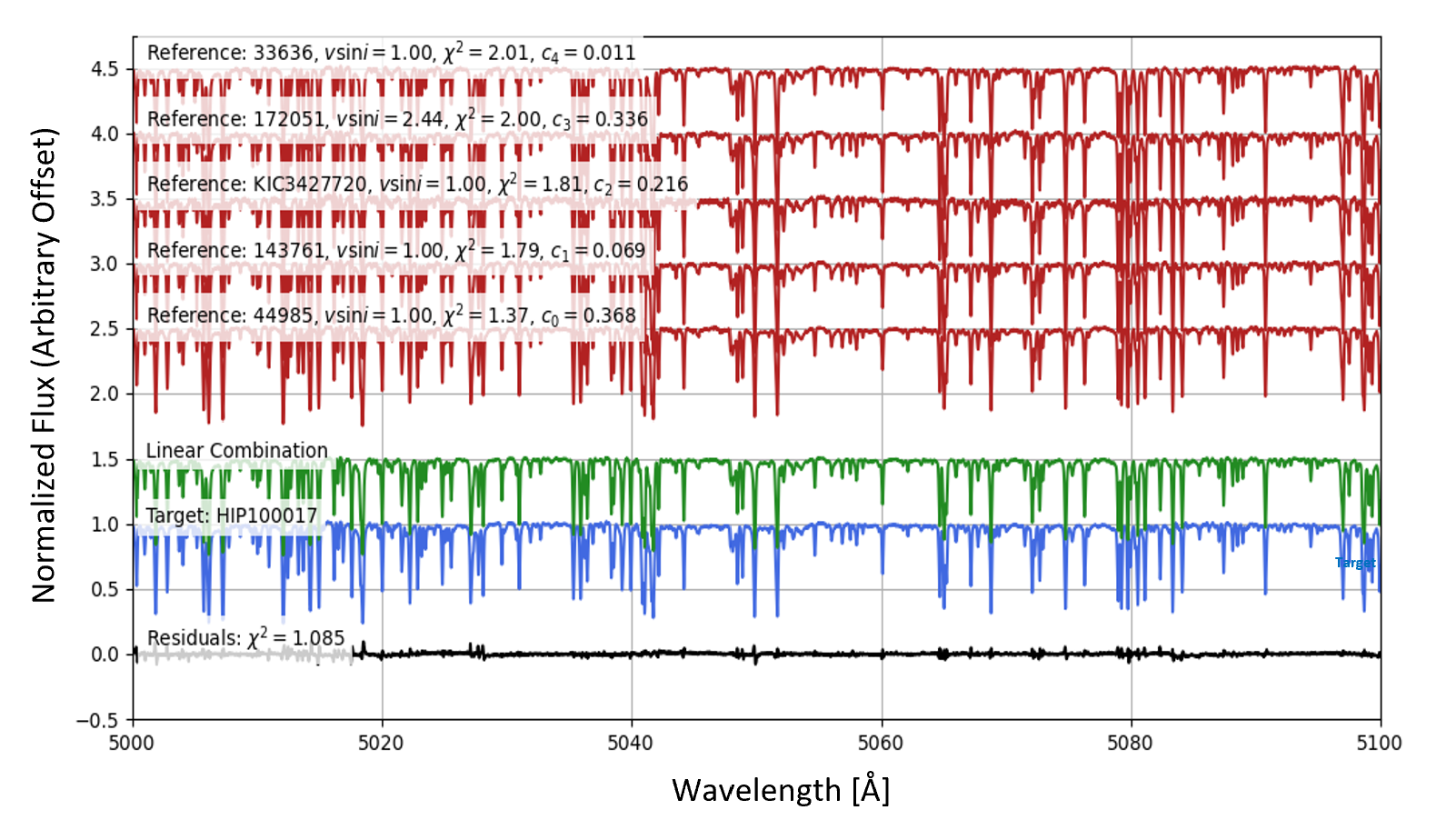}
\caption{An example target spectrum (HIP\,100017), with the five best-matching library spectra and their linear combination. The residual between the target and linear combination is also shown. The residual is free from the stellar absorption features which obscure emission features in the target spectrum. Figures of this form are a direct output of the \texttt{SpecMatch-Emp} model.}
\label{fig:ex_spect}
\end{figure*}

\begin{figure*}[ht]
\centering
\includegraphics[width=0.85\textwidth]{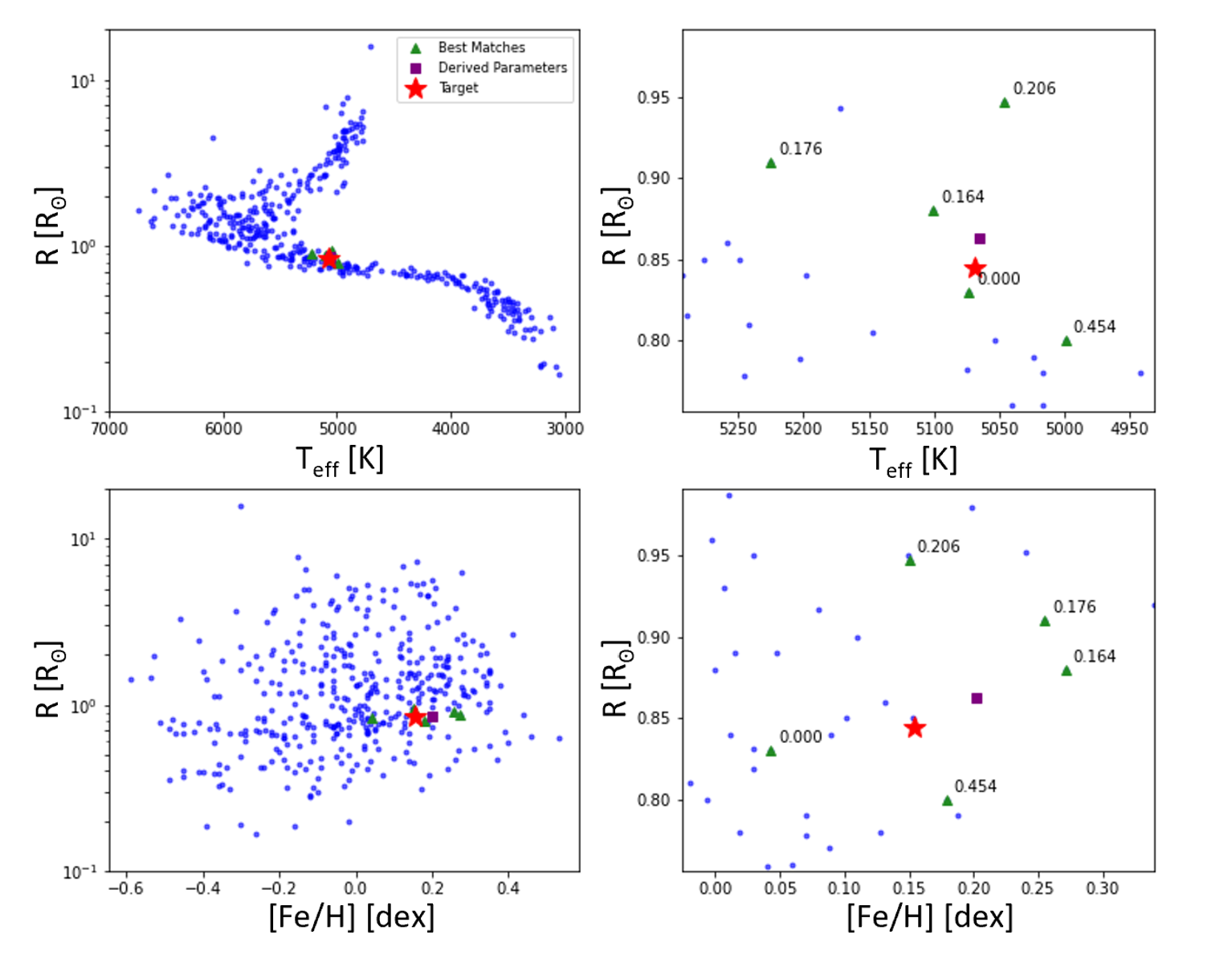}
\caption{Derived stellar properties for an example target star (HIP\,75722, purple square) shown  relative to stars from the \texttt{SpecMatch-Emp} library (blue circles), with the best-matching library stars (green triangles) indicated. Because HIP\,75722 is in the library, we can also compare to its library parameter values (red star). Figures of this form are a direct output of the \texttt{SpecMatch-Emp} model \citep{Yee2017}.}
\label{fig:ex_properties}
\end{figure*}

\begin{figure*}[ht]
\centering
\includegraphics[width=0.75\textwidth]{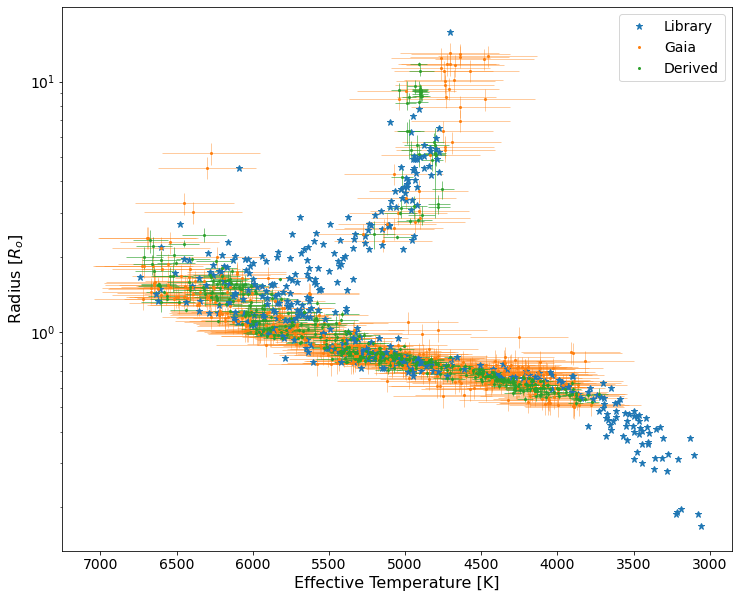}
\caption{Stellar properties derived from APF spectra using the \texttt{SpecMatch-Emp} algorithm and subsequent isochrone analysis for stars within the library bounds (blue star shapes), compared to the properties for the same stars as reported in the Gaia catalog (green crosses). The library stars are also shown for reference (orange crosses). The matching algorithm inherently pulls the derived properties towards the library stars because all matches must come from within the library. The lack of stars cooler than $T_{eff} \approx$ 4800 K reflects the magnitude limit of the APF.}
\label{fig:HR}
\end{figure*}

We use the \texttt{SpecMatch-Emp} model to create a linear combination of the closest matching library spectra to each target, and subtract that model from the target. The resulting residuals provide the basis of our laser search. The \texttt{SpecMatch-Emp} algorithm takes an APF stellar spectrum as input, and shifts it onto the target's rest frame by cross-correlating the target spectrum with a reference spectrum of the same spectral type to determine the relative velocity of the target as observed from Earth. The best reference for shifting is chosen by cross-correlating the target against several reference stellar types to determine the best-matching spectrum. The wavelength scale onto which each target is shifted is constant in $\Delta({\rm log} \lambda)$ such that a given relative velocity produces a constant shift in pixels across the spectrum. The target spectrum is then compared to the \texttt{SpecMatch-Emp} library of 404 stellar spectra corresponding to stars of a wide range of spectral types (from M5 to F1). The algorithm makes a linear combination of the five closest matching spectra using weights determined by performing a least-squares fit to the target, which minimizes the chi-square value between the matches and the target. A rotational broadening kernel is applied to library spectra during the matching process in order to fit for differing projected rotation velocity, \vsini, values between the target and library stars. An example target spectrum and linear combination is shown in Figure \ref{fig:ex_spect}. The library properties of the best-matching stars are weighted by the coefficients of the linear combination, to produce a set of derived properties for the target star. The library contains three empirically determined stellar properties ($R_{star}$,  $T_{eff}$,  and  $[Fe/H]$) for each star, derived using interferometry, asteroseismology, LTE spectral synthesis, and spectrophotometry. Three additional properties (${\rm log}(g)$, $M_{*}$ and $age$) are calculated analytically in the library using isochrone analysis. The derived properties for the example star HIP\,75722, compared to the library properties, are shown in Figure~\ref{fig:ex_properties}. We consider the set of properties derived directly from \texttt{SpecMatch-Emp} an intermediate result and perform the additional step of isochrone analysis to derive a more accurate set of properties as described in Section \ref{subsec:isochrone} (Figure \ref{fig:HR}). The \texttt{SpecMatch-Emp} model, catalog stars, and calibration using Keck HIRES observations are described fully by \cite{Yee2017}.

The primary purpose of \texttt{SpecMatch-Emp} is to derive stellar properties for a target star based on its spectrum. For our laser search algorithm, we also use the \texttt{SpecMatch-Emp} algorithm to produce a residual spectrum by subtracting the linear combination of best-matching spectra from the target spectrum. Our laser search is conducted on the residual spectrum rather than the original spectrum. The residual enables a more sensitive search because it is free from stellar absorption features which can limit the sensitivity to emission lines which fall within absorption features. We quantify our improved sensitivity in Section \ref{subsec:sensitivityadvantage}. 

\subsubsection{Calibration}
\label{subsec:calibration}

\begin{figure*}[ht]
\centering
\includegraphics[width=\textwidth]{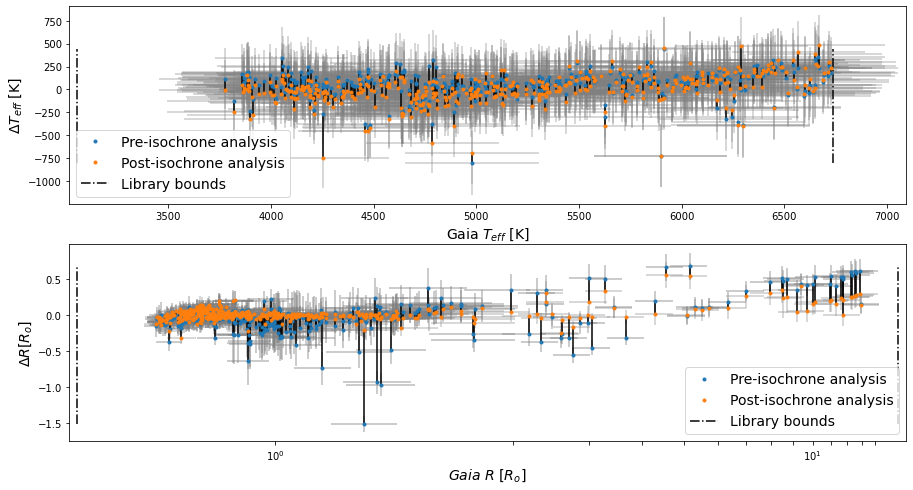}
\caption{Difference between Gaia catalog values and derived values for stars whose Gaia values are within the \texttt{SpecMatch-Emp} library bounds in effective temperature and radius. The sparsity of the library at high radii is shown in the decreased accuracy at high radius values. Applying isochrone analysis significantly improves the results relative to Gaia, especially at high radius values.}
\label{fig:gaia_comparison}
\end{figure*}

To verify and calibrate the APF stellar property results, we compare our results from \texttt{SpecMatch-Emp} to empirical values from the Gaia catalog. For the calibration sample we use the subset of CPS targets observed by the APF which have entries in the \texttt{SpecMatch-Emp} library, 101 stars. To calibrate, we choose the highest SNR observation for each star, removing each star used in the calibration from the library prior to running \texttt{SpecMatch-Emp} on its spectrum. The derived properties can then be compared to the library values to calibrate the stellar parameter determination for APF stars. \cite{Yee2017} validate properties for the library stars in this way using the \texttt{SpecMatch-Emp}  results from Keck spectra in comparison to the library values. Some scatter is expected because the library properties are not known exactly and the matching process is not perfect. The primary cause of systematic errors is the finite span of the library and non-uniform distribution of library stars within that parameter space. Similar to \cite{Yee2017}, we correct systematic trends in the final parameters by assessing differences between the library and derived parameters (as shown in Figure \ref{fig:gaia_comparison}). We identify and exclude target stars whose Gaia-reported properties fall outside of the region spanned by the library, as these cannot be matched well.  In particular, library stars are sparse near the edges of the library parameter region, resulting in a regression towards the mean effect even for stars within the library parameter region. We conduct a similar analysis to that of \cite{Yee2017} for Keck spectra and use our calibration target set in order to verify that the \texttt{SpecMatch-Emp} algorithm performs well for APF observations.

We calculate the difference between the library and derived values for each star in our calibration subset, then fit these residuals as a function of the derived values. The trends are removed from the derived stellar parameters to correct for systematic errors. In choosing the trend functions for each stellar parameter, we  prefer the simplest function, with the fewest free parameters, which significantly reduces the scatter in the residuals between the library and derived values. For effective temperature, we find that a constant offset accounts for most of the systematic error. For metallicity, there is a systematic offset such that low metallicity is overestimated and high metallicity is underestimated requiring a linear trend function that corrects for this systematic error. For radius, we detrend the fractional differences and find that a linear trend best accounts for the systematic trends without introducing unnecessary complexity. Only  the region $1.0  R_\sun < R_{*} < 2.0  R_{\sun}$ requires detrending because the residuals in the region $R_* < 1.0  R_{\sun}$ are essentially flat, and beyond $ R_* = 2.0  R_{\sun}$ the data are very sparse. For metallicity, stellar radius, surface gravity and stellar mass, the detrending is performed only in the regions -1 dex $\leq [Fe/H] < 1$ dex, $1 R_\sun \leq$ $R_{*}$ $< 2 R_{\sun}$, $3 \leq {\rm log}(g) < 6$, and  $ 0.2 M_{\sun} \leq$ $M_{*}$  $< 3 M_{\sun}$, respectively, because the data are too sparse beyond those regions to perform a fit confidently. For $T_{eff}$ and age, a constant offset is used across the full range of values.

\subsubsection{Isochrone Analysis}
\label{subsec:isochrone}
We improve our stellar property results and extend our analysis using the stellar isochrones package \texttt{isoclassify}  \citep{Huber2017,Travis2020} to determine the parameters ${\rm log}(g)$, $M_{*}$, and age. Although \texttt{SpecMatch-Emp} derives six stellar properties, only the library values of $T_{eff}$, $R_{*}$, and $[Fe/H]$ are directly determined in the \texttt{SpecMatch-Emp} spectral library. Isochrones are theoretical lines on the Hertzsprung-Russell diagram representing stars with the same age but different initial masses. Given stellar parallax, photometry, and \texttt{SpecMatch-Emp} initial parameter values for $T_{eff}$, $[Fe/H]$, ${\rm log}(g)$, and $R_{*}$, the  \texttt{isoclassify} model determines the most consistent isochrone providing the stellar age. The location on the isochrone determines the initial mass. In this way a final set of parameters can be probabilistically determined. 

Even for $T_{eff}$, $R_{*}$, and $[Fe/H]$, which are well-defined by \texttt{SpecMatch-Emp}, the isochrone analysis step improves the final derived stellar parameters relative to the Gaia catalog values for our target set (Table \ref{tab:RMS-table}). The RMS between isochrone derived parameters and Gaia stellar parameters decreases after isochrone analysis is implemented, particularly at large radius values. We report \texttt{isoclassify} values for all derived stellar properties in Appendix \ref{subsec:app-isochrone}.

\begin{adjustwidth}{-1in}{-1in}
\begin{table}[]
\setlength{\tabcolsep}{3pt}
\centering
\begin{tabular}{c@{\hspace{-2pt}}ccccccc}
\hline
                                         & \multicolumn{3}{c}{APF SM-Emp to Library RMS}                                          & \multicolumn{4}{c}{APF SM-Emp to Gaia RMS}                                                                                                        \\
\multicolumn{1}{l}{}                     & \multicolumn{1}{l}{}     & \multicolumn{1}{l}{}     & \multicolumn{1}{l}{}             & \multicolumn{2}{c}{Pre-isochrone analysis}          & \multicolumn{2}{c}{Post-isochrone analysis}                                                 \\ \hline
                                         & $\sigma (\Delta T_{eff})$[K]  & $\sigma (\Delta R_*/R_*) [\%]$ & $\sigma ([Fe/H]) [dex]$ & $\sigma (\Delta T_{eff})$[K] & $\sigma (\Delta R_*/R_*) [\%]$ & \multicolumn{1}{l}{$\sigma (\Delta T_{eff})$[K]} & \multicolumn{1}{l}{$\sigma (\Delta R_*/R_*) [\%]$} \\
All targets                              & 113.98                   & 12.53                    & 0.08                             & 159.31                   & 128.64                   & 159.80                                       & 59.91                                        \\
$T_{eff}<4500$\,K                    & 71.92                    & 9.21                     & 0.11                             & 133.97                   & 11.42                    & 137.30                                       & 8.86                                         \\
$T_{eff}\geq4500$\,K            & 125.31                   & 13.49                    & 0.07                             & 167.20                   & 24.29                    & 166.36                                       & 9.19                                         \\
$R\leq1.0R_o$                        & 89.26                    & 8.58                     & 0.08                             & 123.60                   & 11.23                    & 115.00                                       & 6.26                                         \\
$1.0R_o<R\leq2.5R_o$  & 144.84                   & 17.01                    & 0.08                             & 204.40                   & 27.38                    & 207.66                                       & 5.74                                         \\
$2.5R_o<R\leq6.0R_o$ & \multicolumn{1}{l}{}     & \multicolumn{1}{l}{}     & \multicolumn{1}{l}{}             & 175.90                   & 36.91                    & 185.69                                       & 22.68                                        \\
$R>6.0R_o$                     & \multicolumn{1}{l}{}     & \multicolumn{1}{l}{}     & \multicolumn{1}{l}{}             & 232.55                   & 49.06                    & 260.62                                       & 23.37                                        \\ \hline
\end{tabular}
\caption{RMS values for detrended derived APF stellar property values (before and after isochrone analysis), compared against the \cite{Yee2017} library values, and compared against the Gaia catalog values. The latter is determined only using stars whose Gaia properties are within the $T_{eff}$ and radius bounds of the library. It is especially apparent in the Gaia RMS values that restricting our targets to the region $T_{eff}$ \textless 4500 K and R \textless 1.0 $R_\sun$ greatly improves the RMS values. Note that the library was constructed to contain only well-behaved stars and thus the RMS is expected to be lower for the calibration set compared to library values than for the overall dataset compared to Gaia values.}
\label{tab:RMS-table}
\end{table}
\end{adjustwidth}

\subsection{Laser Search} \label{subsec:search}

Next, we describe our laser search method, which we subsequently carry out on both our residuals and spectra. We search the spectra in order to have a baseline for evaluating our residual search.

The first step in our search for laser emission lines is to impose a threshold in flux. We then assess threshold crossing events with a series of criteria that ultimately eliminates all of our events as potential laser emission with a non-astrophysical origin beyond the Earth. Our intensity threshold is determined by calculating the typical noise in each target spectrum and choosing some number of noise intervals above the median flux value, such as ten (less sensitive) or three (more sensitive). A more sensitive search flags more threshold crossing events, but classifying those events is more challenging as their number increases to include a greater number of false positives, such as cosmic rays and night sky emission lines. We will show in Section \ref{subsec:persistence} that a SETI-focused observing practice provides the most sensitive search possible, while searching residuals instead of the spectra themselves adds a small additional sensitivity to the search (Section \ref{subsec:sensitivityadvantage}). Next, we define our event threshold, and discuss identification and classification of events.

\subsubsection{Threshold Crossing Criteria}
\label{subsec:thresholdcrossing}

\begin{figure}[H]
\begin{center}
    \includegraphics[width=8cm]{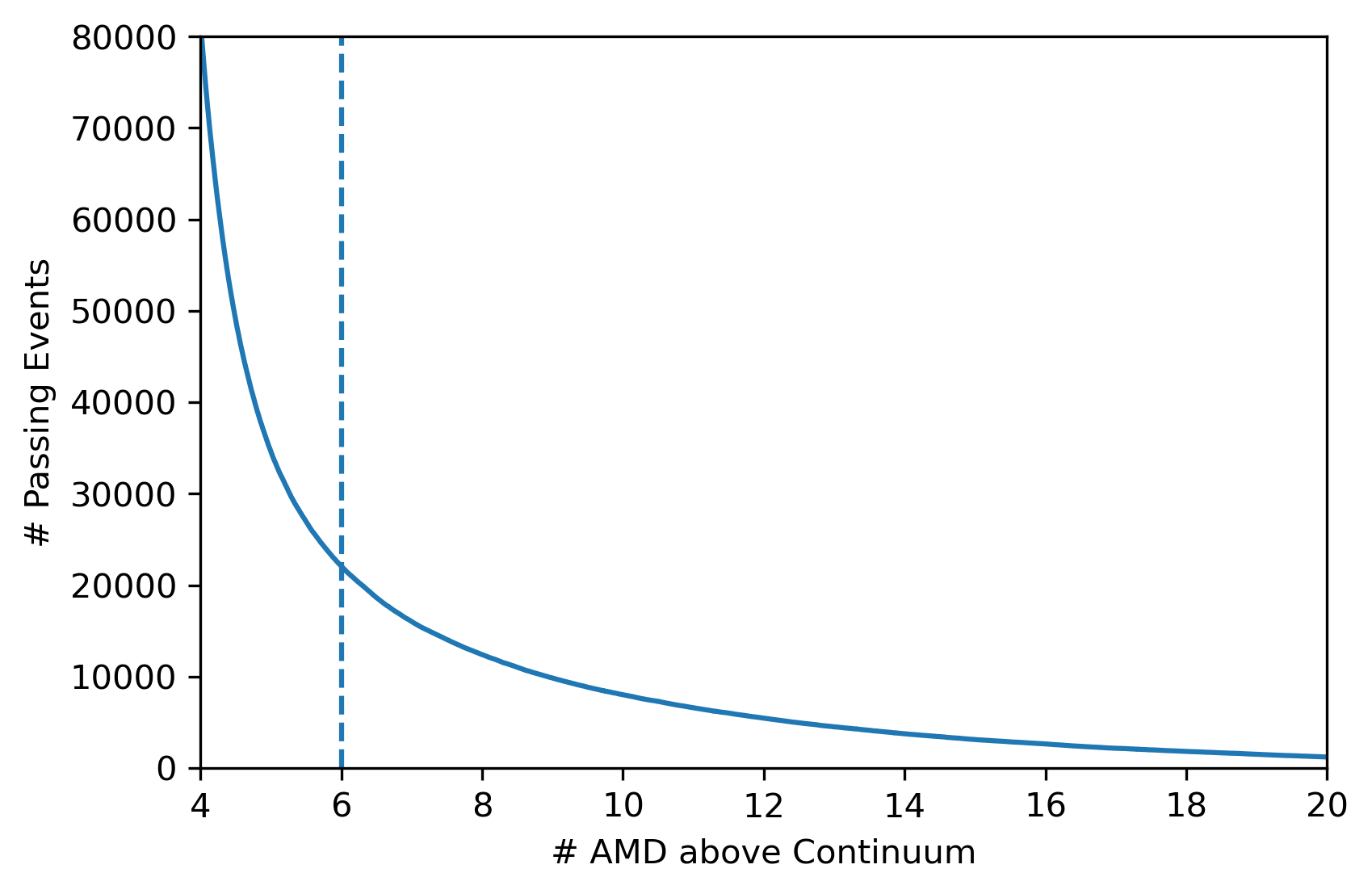}
    \caption{The number of events we obtain with different thresholds for the residual search. These are threshold crossing events that pass the Gaussian profiling test described in Sections~\ref{subsec:thresholdcrossing} and \ref{subsec:gaussianprofile}. We chose a threshold of six times the AMD of the residuals to maximize sensitivity while minimizing the number of false positives, which increases as the threshold is lowered.
    }
    \label{fig:determiningthreshold}
    \end{center}
\end{figure}

\begin{figure}[H]
\begin{center}
    \includegraphics[width=8cm]{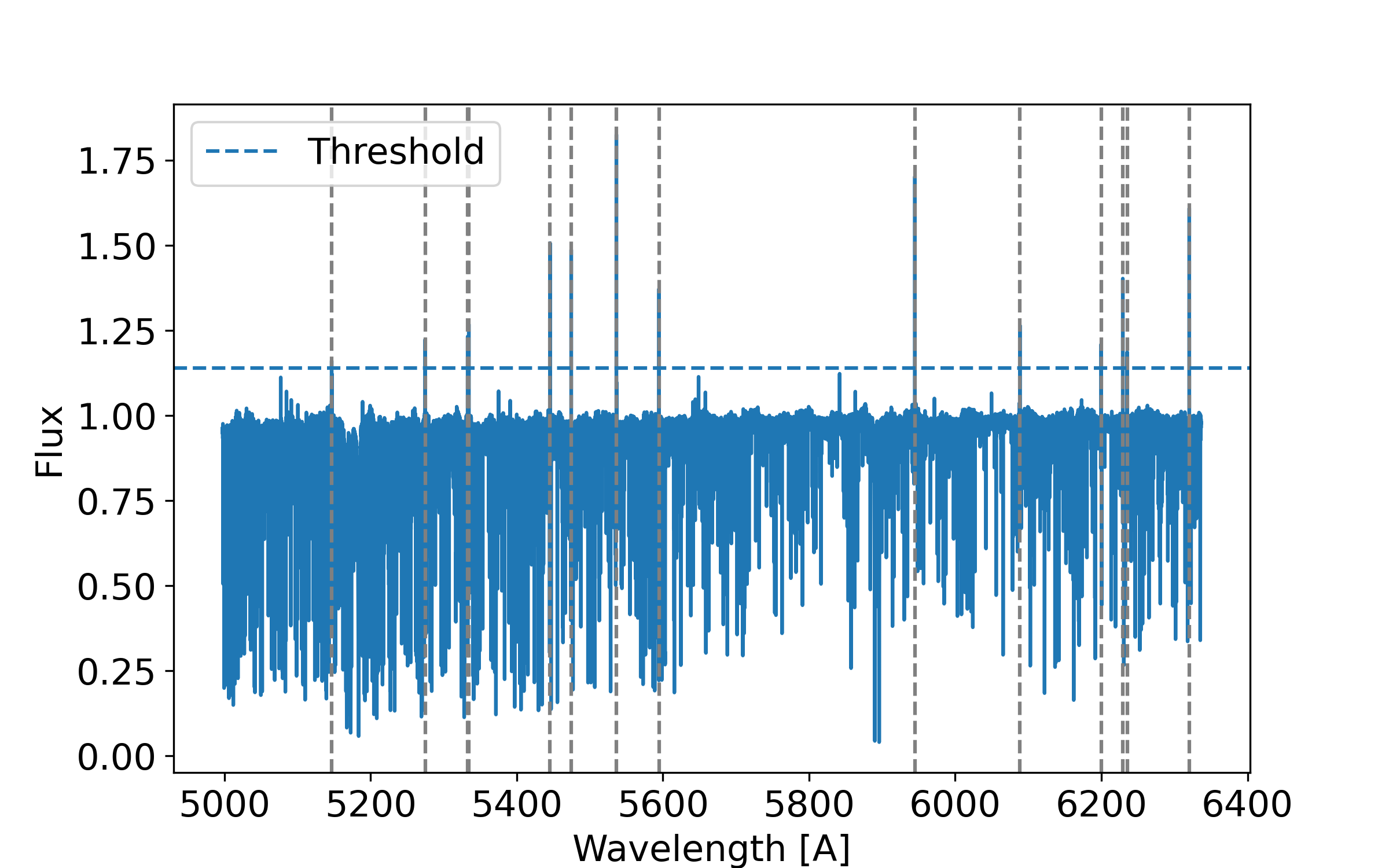}
    \caption{Events must rise above the threshold of the spectrum or residual. The vertical dashed lines mark each threshold crossing event. The horizontal line at roughly 1.1, for this spectrum, marks the threshold at six AMD of the residual of this spectrum. The wavelength coverage of the \texttt{SpecMatch-Emp} library, from 5000 to 6300\,\AA\ is shown. This is a subset of the full coverage of the APF-Levy spectrograph. }
    \label{fig:AboveThreshold}
    \end{center}
\end{figure}

When defining the threshold, we choose the absolute median deviation (AMD) of the residual as our noise metric, following \cite{TellisMarcy2017}, who chose a value of six AMD above the continuum of an individual order as their threshold. We calculate and utilize the AMD of the residuals when searching both spectra and residuals because a residual is the spectrum minus a model of the stellar continuum. There is no multiplication or division involved; therefore, the deviation due solely to noise should be the same in both spectra and residuals. For the spectra, the AMD is an overestimate of the noise because it captures both variations due to noise and those due to absorption features. We want a value for sigma that is only dependent on noise and not absorption features. We choose a threshold of six times the AMD of the residuals, balancing the number of false positive threshold crossing events to vet with maximizing the sensitivity of our laser search algorithm (see Figure \ref{fig:determiningthreshold}). In Figure \ref{fig:AboveThreshold}, an example \texttt{SpecMatch-Emp} processed spectrum with the threshold value overplotted highlights threshold crossing events and shows that the threshold is very close to the continuum, indicating a sensitive search. Using this method allows us to quantify the improvement in the sensitivity when searching residuals compared to searching spectra directly for laser lines (section \ref{subsec:injection_recovery}).

\subsubsection{Vetting Criteria}

We define a threshold crossing event, as a group of consecutive pixels that rise above the threshold of a spectrum or residual. We establish the following criteria to vet these events. They must be Gaussian-shaped with a FWHM greater than the point-spread function (PSF, also known as the line-spread function) of the APF. Each event must be persistent across all observations of the star taken on a single night, and must not be at a wavelength of a known night sky emission line. Our laser detection algorithm imposes these tests on all target spectra and residuals. The residuals yield a higher sensitivity than the spectra because they do not have absorption features that could hide or alter the shape of an event. 

Some previous works have also required all photons to fall onto the stellar trace on the 2D spectrum, indicating a source is coincident with the target star. An event that passes through the optics of the telescope will fall completely within a single order, whereas a cosmic ray hit can fall inside or outside the order containing the stellar light. Because we require events to repeat across all observations of the target star on a given night, we are able to successfully remove false positives due to cosmic ray hits without inspecting the 2D spectra. Therefore, we do not need to implement this criterion. Other works which require that light fall on the stellar trace limit their spatial search sensitivity. As a reference for the magnitude of this effect, at the median distance of a star in our target set (78.47 ly) this criterion would prevent detection of laser emission from greater than 60AU from a target star in the spatial direction and 12AU in the dispersion direction due to the 1\arcsec~by 5\arcsec~size of the Levy decker.

\subsubsection{Gaussian Profile Assessment}
\label{subsec:gaussianprofile}

We require that events are Gaussian-shaped to indicate that they have been convolved with the PSF of the telescope and therefore have passed through the optical path. We fit a Gaussian to each event and calculate the Mean Squared Error (MSE) to measure how closely the event fits a Gaussian. We first identify the edges of the event in the wavelength direction as the two points where the slope is zero. We choose this method as opposed to choosing the wavelengths at which the event falls below the continuum in order to be sensitive to events that fall into absorption features. We then fit a Gaussian to the event and calculate the MSE. If the MSE is under our prescribed limit, the event passes onto the next test. 

\begin{figure}[H]
\begin{center}
    \includegraphics[width=8cm]{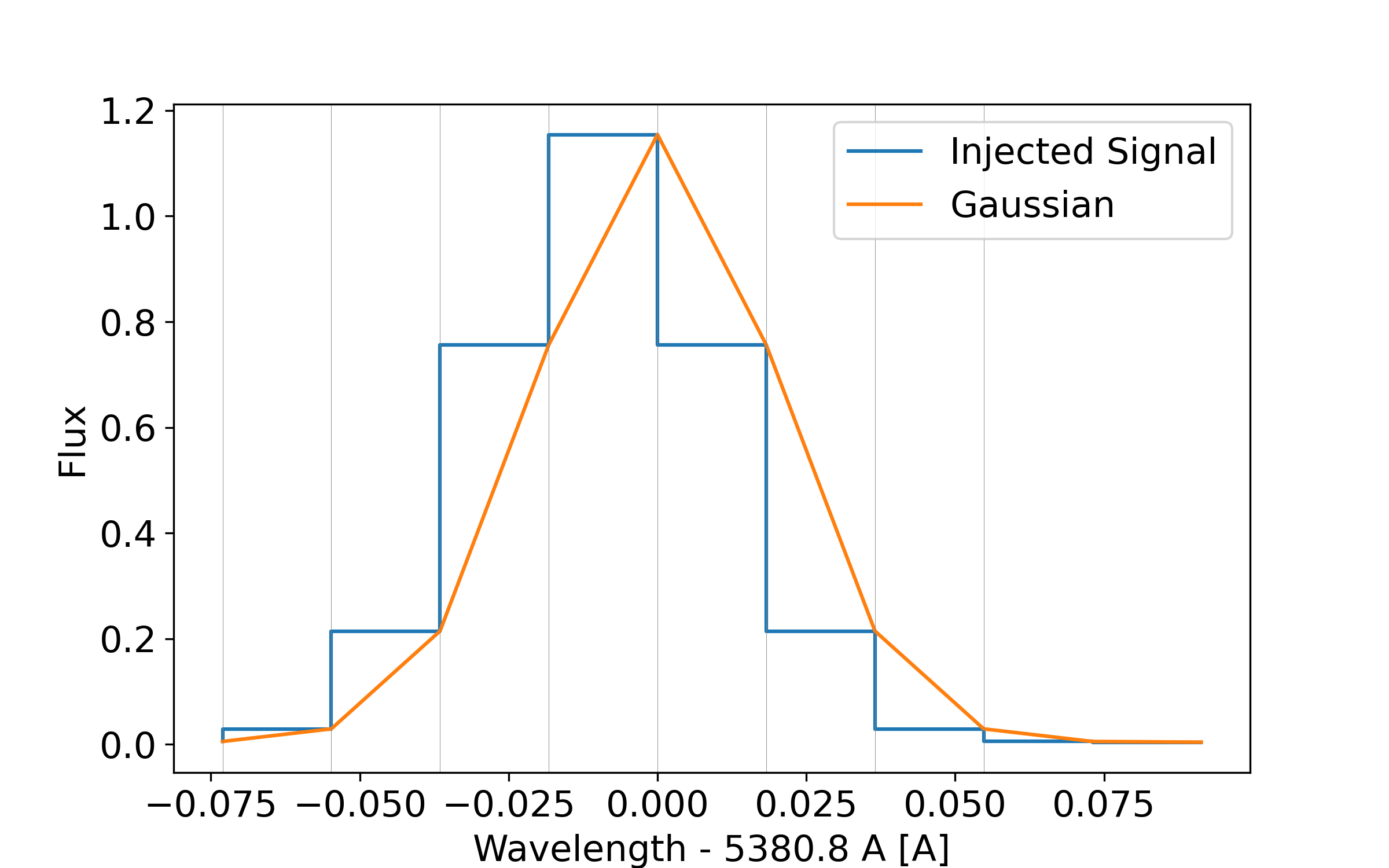}
    \caption{An example of a signal artificially injected into an APF spectrum (blue) with the Gaussian model overplotted (orange). The value of the MSE for this example is 0.018.}
    \label{fig:injected_signal}
    \end{center}
\end{figure}

To determine what MSE to set as the limit for determining whether or not the event is Gaussian-like, we conduct an injection and recovery procedure. We inject 50 artificial signals each into 500 spectra, resulting in 25000 total artificial injections, and pass the spectra with injected signals through our detection algorithm. For all the injected signals that are recovered, we record the MSE of the signal against a best fit Gaussian. An example of an injected signal is shown in Figure \ref{fig:injected_signal}. Ninety-five percent of the artificially injected signals have MSE values under 0.87, however, the MSE values reach a maximum of 2.1. These outliers represent signals that fall into stellar absorption features, which can affect the shape of a laser signal, causing it to lose its Gaussian shape. We used the maximum MSE of 2.1 as our MSE limit because we want to recover every artificial injection, including those that fall into stellar absorption features. Because residuals reduce the effect of absorption features, the residual search yields increased sensitivity to signals that fall onto these absorption features. 

\begin{figure}[H]
\begin{center}
    \includegraphics[width=8cm]{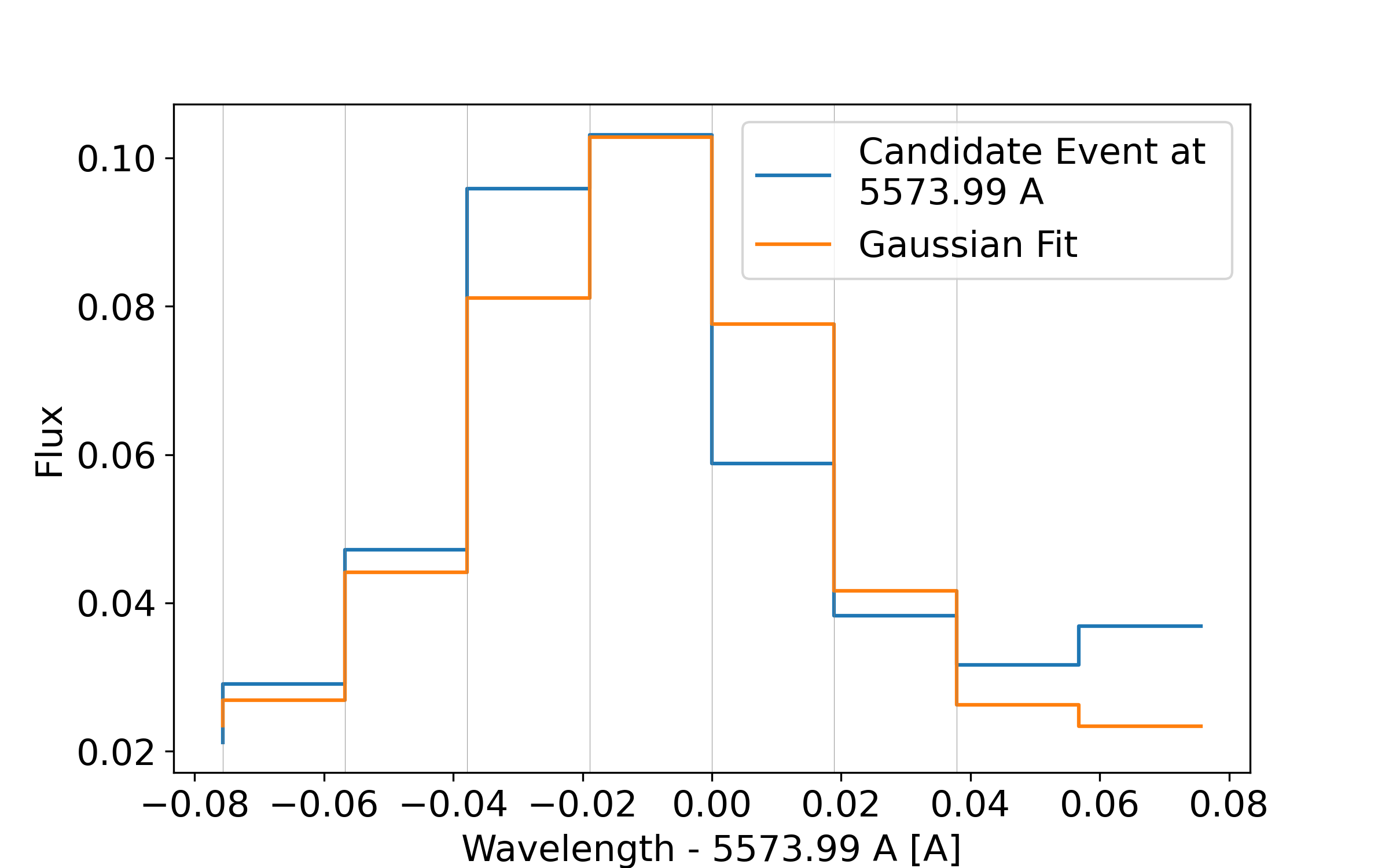}
    \caption{An example of an event that passes the Gaussian profile test with an MSE of 0.2 and a FWHM of 3.4 pixels.}
    \label{fig:gaussian_vetting_candidate}
    \end{center}
\end{figure}

The Gaussian modeling of each event offers two vetting methods. The first is the MSE limit and the second is a check that the event FWHM must be wider than 2.7 pixels, which we determine to be the lower limit on the APF PSF by examining pinhole decker Thorium-Argon lamp spectra. We choose the pinhole decker instead of the science decker as a conservative lower limit on the FWHM of the PSF. Events with profiles narrower than 2.7 pixels are due to cosmic rays or pixel flaws.  If the event is wider than the PSF lower limit, it moves on to the next test. Figure \ref{fig:gaussian_vetting_candidate} shows an example of an event that is categorized as Gaussian-like and is wider than the PSF of the APF.

\subsubsection{Exploring Night Sky Emission Lines}
\label{subsec:nightskyemissionlines}

Next we check for coincidence of our events with night sky (airglow) emission lines, as they match the profiles of the potential laser lines in our search. To understand the magnitude of this issue, we search a subset of our APF target spectra for known emission lines cataloged by \cite{slanger2003} using the Hamilton Spectrograph at Lick Observatory. This night sky emission line catalog contains 266 identified emission lines over Mt. Hamilton, covering a wavelength region $3800 - 9200$\,\AA\ with a spectral resolution of 45000 per pixel. The \citeauthor{slanger2003} catalog was created with data collected in 1999 October using spectra taken at three different positions in the sky: southwest over San Jose, on the meridian, and southeast away from San Jose. The two latter regions were used to differentiate between night sky lines from San Jose such as sodium lines from streetlights and natural night sky lines like those emitted from OH or oxygen. Each of the three regions had exposures of 45 minutes. The region over San Jose received a total of 3 hours, and the other two regions received a total of 2 hours and 15 minutes each.

As a test case, we  search for known sky emission lines in 60 observations of Boyajian’s Star (KIC\,8462852). This star is amenable to this analysis because it has a high rotational velocity, which broadens the stellar absorption features, allowing us to analyze the spectra without worrying about various absorption features that could interfere with the emission lines. Also, all observations of Boyajian’s Star have exposure times of 30 minutes, the longest among all our APF observations, resulting in the strongest night sky emission features among all our stars. We first co-add all 60 spectra in the observatory rest frame, eliminating cosmic ray hits by taking the 99th percentile of every pixel across all 60 spectra. The nearly featureless nature of the spectra allows co-adding in the observatory frame, which is typically not feasible. We then overplot the 266 identified lines from the \cite{slanger2003} catalog and record all events (consecutive pixels) that have a flux value higher than the threshold at the location of a known emission line. We allow the location of the emission line to vary by one pixel, a reasonable amount of variation over one night, to account for positional changes as the instrument is not mechanically nor thermally stabilized. For every recorded event, we overplot each of the 60 individual spectra at that wavelength and record all the spectra that have a flux value higher than the continuum at the known emission line.

\begin{figure}[H]
\begin{center}
    \includegraphics[width=10cm]{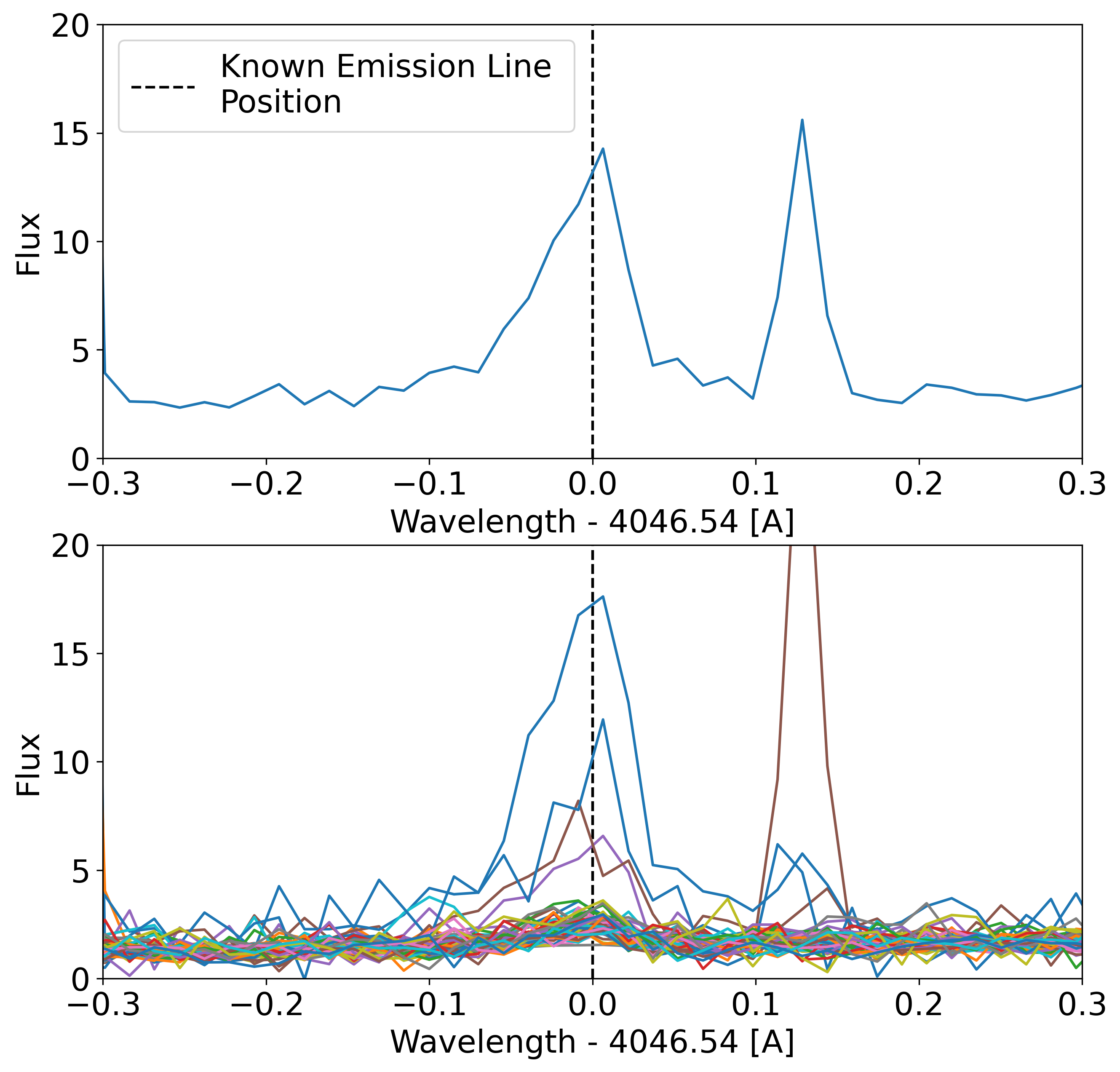}
    \caption{The combined spectra of 60 Boyajian's Star observations (99th percentile of every pixel) show an emission line at 4046.54\,\AA, a known line from the \citeauthor{slanger2003} catalog (Top). The line to the right could be a sky emission line or a cosmic ray hit, but to distinguish between them requires examination of the individual spectra (Bottom). Each observation of Boyajian's Star has an emission line at 4046.54\,\AA, but only four rise above the continuum level. The emission line to the right of the known emission line is likely a cosmic ray hit because it is largely due to one observation.}
    \label{fig:boyajianstar}
    \end{center}
\end{figure}

Out of the 266 emission lines from the \citeauthor{slanger2003} catalog, only one of the lines was found in our Boyajian’s Star data, and for that one emission line, only four out of our 60 spectra showed a significant emission line, which we define as at least three consecutive pixels rising above the continuum of the spectrum (Figure \ref{fig:boyajianstar}). This night sky emission line is only present in 4 out of 60 of the Boyajian's Star spectra, suggesting that the emission line will appear weakly or not at all in other APF observations. Sky emission lines are dependent on orientation and observation time, so a one-time catalog can be helpful but will not be exhaustive. We mark this wavelength to remove candidates that may fall onto this sky emission line, but the remaining 255 cataloged lines were not found in our data, suggesting that the \citeauthor{slanger2003} catalog is insufficient in identifying all our sky emission lines. We impose an additional check for sky emission lines by analyzing events that fall on the same observed wavelength, which is described in Section \ref{subsec:nightskyskyemission}.

\subsubsection{Persistence and Order Proximity Requirement}
\label{subsec:persistence}
To ensure that the events are persistent in time, we require each event to be detected in all of the  observations of the same star in a given night. This test rejects noise which happens to mimic the shape requirements we impose. Rejected events include cosmic rays which hit the detector at an angle, or random noise which happens to mimic a Gaussian shape. Such events would pass the above tests on shape, but would not appear in all consecutive observations. The persistent events then move on to the final vetting.

\section{Results} \label{sec:results}

We searched \NSpectraSearched\ APF spectra, corresponding to \NSearched\ unique stars. We define our final candidates as events that passed the threshold test, Gaussian profiling test, and persistence test. We first identified events using the threshold described in Section~\ref{subsec:thresholdcrossing}, which yielded 12161 threshold crossing events when the spectra themselves are searched and 22247 when the residuals are searched. We then imposed the Gaussian profiling test (see Section~\ref{subsec:gaussianprofile}) to eliminate cosmic ray hits that have a narrower profile than events that pass through the spectrometer, narrowing our pool down to 1472 events in the spectra and 1518 events from the residuals. By requiring events to be persistent throughout all observations of a star in a single night (see Section~\ref{subsec:persistence}), we narrow the pool down to 55 final candidates in 26 stars' residuals and 6 final candidates from 6 stars when searching the spectra. This cut allows us to dramatically reduce false positives algorithmically, demonstrating the importance of taking consecutive observations for a SETI laser search. After this cut, we had 9.2 times more candidates in the residuals than in the spectra. All six of the candidates found in the spectra are also found in the residuals, so we solely analyzed the candidates from the residuals. In the next few sections we manually vet all 55 candidates and exclude each from our final pool as either a night sky emission line or due to a poor \texttt{SpecMatch-Emp} model fit caused by high rotational velocity, low metallicity, or a poorly matching absorption feature.

\subsection{False Positive Scenarios}
\subsubsection{Night Sky Emission Lines} \label{subsec:nightskyskyemission}

In the 55 candidates, there are six that are not unique in the observed frame, meaning there is another star with a candidate at the same observed wavelength (to the nearest pixel). If there are multiple candidates across different stars that fall onto the same observed wavelength, the candidate is likely a night sky emission line. The corollary to this in a radio SETI search is an event with zero drift rate that is due to an emitter on Earth. To further explore this, we ran the laser search on all the spectra and residuals at an increased sensitivity, using three AMD above the continuum as our threshold instead of six. At a higher sensitivity, we expect to see more false positives due to weak night sky emission lines. This higher sensitivity search allows us to be sensitive to sky emission lines in our data that we missed through the \citeauthor{slanger2003} catalog search described in Section \ref{subsec:nightskyemissionlines}. By identifying the observed wavelengths that repeat in the candidates using a threshold of three AMD, we effectively mask out these lines in our six AMD search. The three AMD run yields 24 final candidates in the spectra and 1677 final candidates in the residuals. In the residuals, there are 177 wavelengths in the observed frame that have three or more candidates from different stars. Of these 177 wavelengths that have repeated candidates in the three AMD search, four of them are present in the six AMD run, corresponding to 11 candidates in our final pool. 

\begin{figure}[H]
\begin{center}
    \includegraphics[width=8cm]{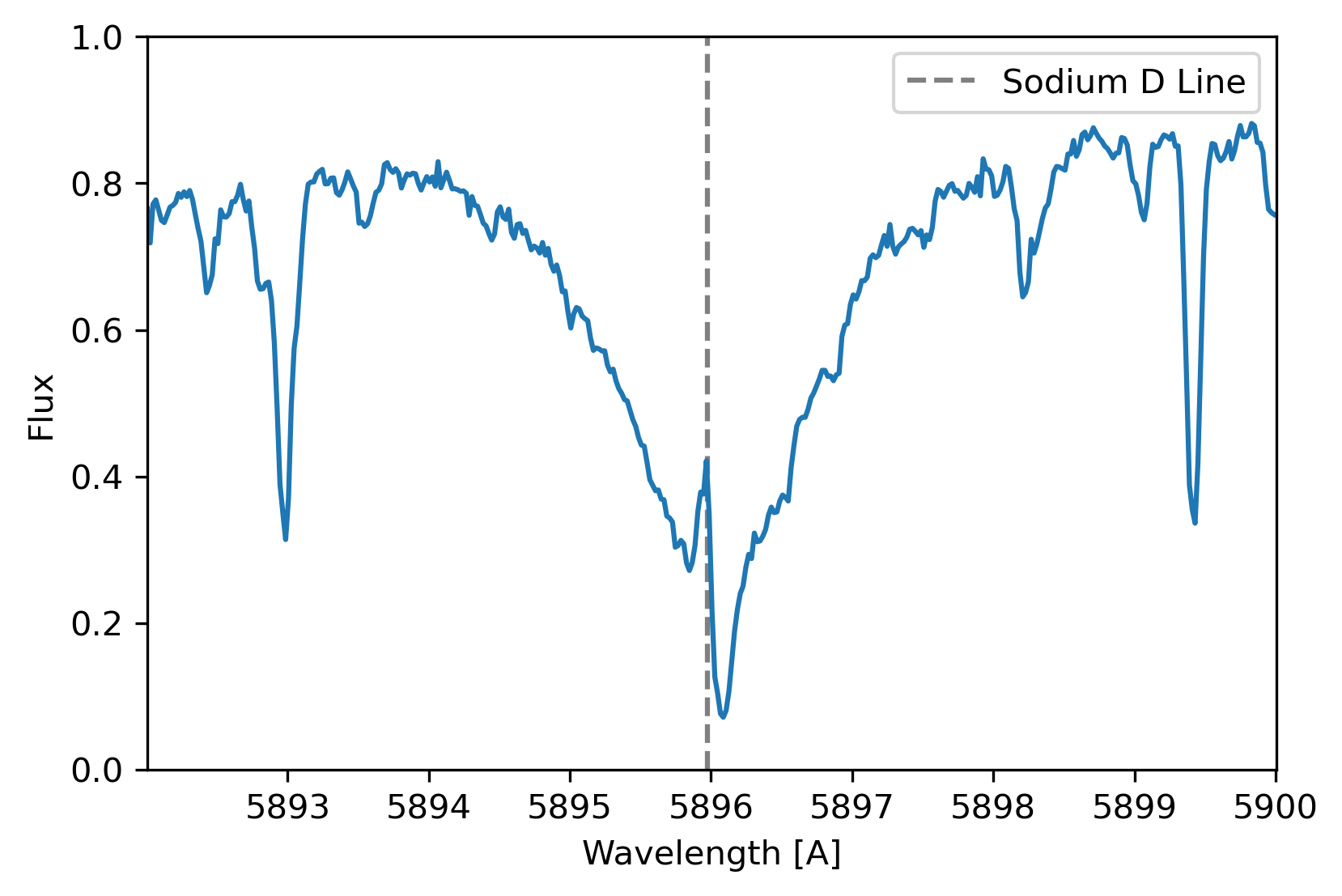}
    \caption{An example of a sodium D night sky emission line that was removed from the final pool of candidates. This emission feature (due to nearby San Jose, CA streetlight contamination) is offset from the core of the stellar absorption line due to the stellar systemic velocity relative to the Earth and the Earth's motion relative to the barycenter of the solar system.}
    \label{fig:sodiumd}
    \end{center}
\end{figure}

We found four night sky emission lines in our final pool of candidates, located at 5083.50, 5405.88, 5577.41, and 5895.97\,\AA. These emission lines showed up in 11 candidates from ten different stars: HIP\,68030, HIP\,47990, TIC\,142276270, HIP\,101262, HIP\,98677, TIC\,198456933, TIC\,219778329, HIP\,45836, HIP\,114430, and HIP\,117463. None of these lines are included in the \cite{slanger2003} catalog, which confirms that the catalog does not contain all our sky emission lines and necessitates this additional check. Figure~\ref{fig:sodiumd} shows an example of a well understood night sky emission line,  a known false positive due to street light contamination from nearby San Jose, CA. The figure also shows our ability to retrieve emission features that fall into stellar absorption lines.

By classifying false positives due to night sky emission lines, we eliminate 11 out of our 55 candidates, leaving 44 to vet.

\subsubsection{ Targets with High Rotational Velocity} \label{subsubsec:highvsini}

Six of our targets have projected rotation velocity, \vsini, values reported to be greater than 10\,km s$^{-1}$ by the \texttt{SpecMatch-Syn} model \citep{Petigura2017}, where $i$ is the angle of inclination of the star's rotation axis to our line of sight. These spectra show broad absorption features, preventing \texttt{SpecMatch-Emp} from successfully subtracting a model from the spectra to generate residuals. Though SpecMatch-Emp does apply a rotational broadening kernel to fit for \vsini~during the matching process, the model is finite in its range of allowed values, and so we expect to encounter some false positives from rapidly rotating stars which we are subsequently able to vet. The matched spectra have deeper and narrower absorption features than these target spectra, resulting in an over subtraction of the spectra. Through classifying false positives due to poor model subtraction from broad absorption features, we eliminate nine out of our 44 remaining candidates, leaving 35 to vet.

\subsubsection{Targets with Low Metallicity} \label{subsec:lowmetallicity}

Very low metallicity stars have shallower absorption features leading to poor matches with stars in the \texttt{SpecMatch-Emp} library,  resulting in false positives in the search of the resulting residual that are due to imprecise subtraction of a library spectrum. This category of false positives is identified through observing an emission line in a residual but only an absorption feature in the corresponding spectrum at a specific wavelength. Two stars in our final pool of candidates are metal-poor with metallicities under -0.3 dex. HIP 7760 has [Fe/H] = -0.42, while HIP 68030 has [Fe/H] = -0.35 leading to 13 and  8 false positives, respectively.  Figure~\ref{fig:smerror} (left panel) shows a wavelength section of HIP\,68030 revealing a threshold crossing event in the residuals at the locations of several absorption features in the spectrum. Through classifying false positives due to poor model subtraction from shallow absorption features, we eliminate 21 out of our 35 remaining candidates, leaving 14 to vet.

\subsubsection{Poor Matches to the \texttt{SpecMatch-Emp} Library} \label{subsec:SMerror}

\begin{figure}[H]
\begin{center}
    \includegraphics[width=14cm]{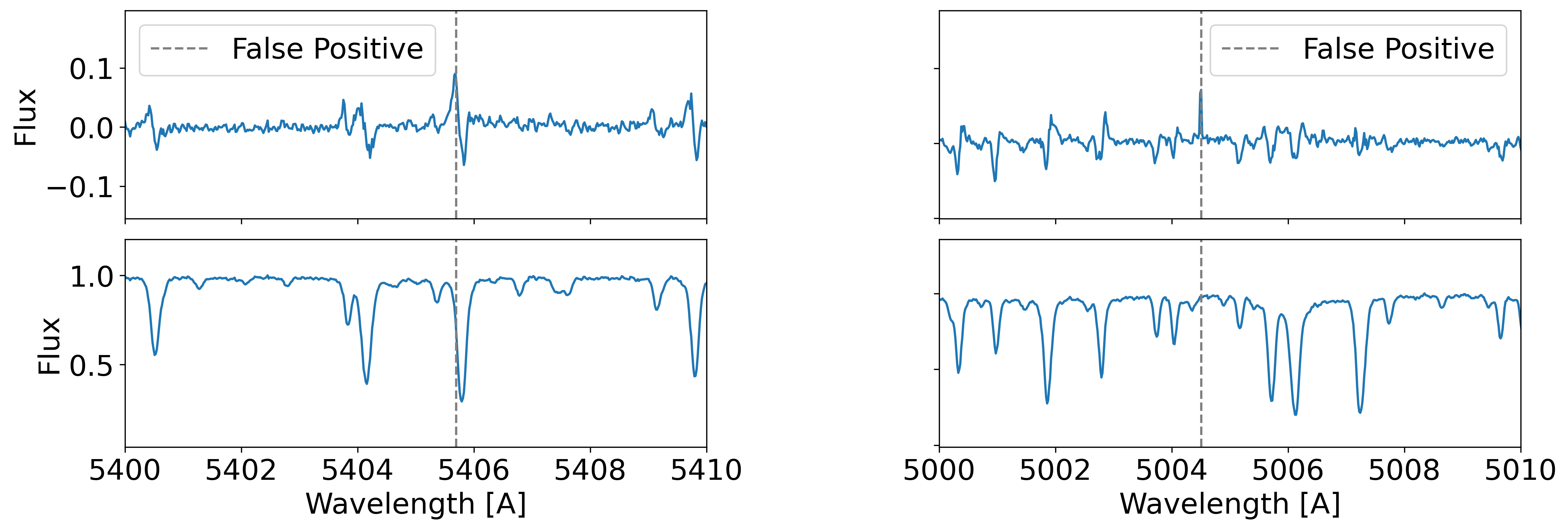}
    \caption{Two examples of false positives in the residuals due to the incorrect subtraction of absorption features. Top: The residual spectrum. Bottom: The corresponding spectrum near the candidate. The left plots show a false positive due to a poor \texttt{SpecMatch-Emp} fit from a low metallicity target, HIP\,68030. This target has a low metallicity of -0.35 dex \citep{Petigura2017}, resulting in shallow absorption features that \texttt{SpecMatch-Emp} cannot fit precisely. The right plots show a false positive due to an absorption feature present in the \texttt{SpecMatch-Emp} library but not present in the target star spectrum. The model incorrectly subtracted out an absorption feature at this wavelength causing an anomalous feature that triggered our laser search algorithm. The alignment of the candidates and absorption features is the visual confirmation used to eliminate these events.}
    \label{fig:smerror}
    \end{center}
\end{figure}

Similar to the continuum subtraction challenges that arise for low metallicity stars, some candidates are due to an imperfect subtraction of absorption features. This can be seen when an emission line is found in the residuals, but only an absorption feature or continuum is found at that wavelength in the spectrum (Figure~\ref{fig:smerror}, right panel). By-eye examination of these poor matches eliminates the remaining 14 candidates.

\subsection{Signal Intensity Retrieval}
\label{subsec:injection_recovery}
There are several factors which limit the sensitivity of our laser search algorithm, both in terms of detecting that a laser emission line exists and recovering the wavelength and intensity of each identified candidate. In order to characterize the accuracy with which the position and intensity of signals can be recovered after processing through the \texttt{SpecMatch-Emp} algorithm, we injected 3600 Gaussian signals of known intensities into APF spectra, verified that these signals persist in the processed spectra and residuals, and quantified the accuracy with which we recover their positions and intensities. 

Six signals are injected into each of 600 target spectra at 5461.81\,\AA, 5490.07\,\AA, 5526.77\,\AA, 5555.37\,\AA, 5593.30\,\AA, and 5622.23\,\AA. We choose these locations to produce injections into both the centers of spectral orders and the edges where adjacent orders have overlapping wavelength coverage. The amplitudes  of each injection are one and a half, two, and three times the local baseline flux for the first two, second two and final two injection wavelengths, respectively.

It is important that we are able to accurately recover the location and intensity of any candidate laser emission. In addition to a Poisson uncertainty on the number of photons detected, errors are introduced in tracing the relative height of a signal (after resampling, deblazing, normalizing, and subtracting the linear combination of best-matching spectra) back to the raw number of photons which produced the peak. We quantify these uncertainties in the following manner. For each target, we save an array of unprocessed raw photon counts for each pixel and resample onto the reference wavelength scale without normalizing or deblazing, in order to preserve the photon statistics. We fit each injected signal for its position and height in the processed spectrum.  We calculate a recovered position by shifting the fit position value by the wavelength interval corresponding to the star's relative velocity, then determine the raw photon count corresponding to the resampled wavelength bin closest to the recovered position. 

We are able to recover the positions of identified peaks to within 0.04\,\AA\ (approximately two pixels) accuracy. The intensity determination is slightly more accurate near the centers of spectral orders compared with the edges, but is recovered with an uncertainty of less than $\pm10 \%$ across the spectrum, and a median error of $\pm \%2.5$.

\subsection{Detection Sensitivity}
\label{subsec:sensetivity}
 We define an approximate detection sensitivity of our laser search algorithm in terms of the luminosity and proximity of a laser source required for detection. With our chosen six AMD threshold on event intensity, we calculate the flux received from the host star necessary for a laser to be flagged by our algorithm. Following \cite{Lipman2019}, we define the number of photons emitted by a laser of  luminosity $L_{em}$, with wavelength $\lambda_l$ at distance $d_E$ from Earth from an emitting telescope of diameter $d_T$ during exposure time per observation $\Delta t$ over $N_{obs}$ observations as

\begin{equation} \label{eq:num_em_laser}
    N_{em} = \frac{L_{em}\Delta t N_{obs}}{E_{ph}} = \frac{\lambda_l L_{em}\Delta t N_{obs}}{hc} 
\end{equation}

\noindent where $E_{ph}$ is the energy per photon. Because we treat observations individually without stacking, in this work $N_{obs} = 1$. The number of photons detected by the APF, given an efficiency of $\epsilon$ and collecting area of $A_{APF}$, is 

\begin{equation} \label{eq:num_det_laser}
    N_{det} = \epsilon N_{em}\frac{4 A_{APF}}{\pi d_{E}^2 \theta^2} = \epsilon\frac{4\Delta t N_{obs} L_{em}A_{APF} d_T^2}{(1.22)^2 \pi hc \lambda_l d_E^2} 
\end{equation}

\noindent where the resolving angle $\theta$ of the emitting telescope obeys $\theta = \frac{1.22\lambda_l}{d_T}$, and the small angle approximation leads to  $d_Esin(\theta/2) \approx d_E \theta/2$. In Section \ref{subsec:search}, we require that a feature reach a specific height above the stellar baseline to be considered a threshold crossing event. This threshold is defined such that 

\begin{equation} \label{eq:requirement}
    N_{det} \geq \alpha (AMD)
\end{equation}

where $\alpha$ is the chosen detection threshold in absolute median deviations (chosen to be $\alpha = 6$). In order to phrase the required  intensity in terms of Equation \ref{eq:num_det_laser}, we calculate the AMD of the residual in units of photons, rather than the relative units used in Section \ref{subsec:search}. We convert the residual into photon units by adding one and then multiplying by the baseline flux of the target star. The baseline flux is determined by taking the median flux value of order 45 which has a center wavelength at 5900\AA. The AMD of this scaled residual then represents the scatter in the residual in units of photons.

The requirement of Equation \ref{eq:requirement} becomes

\begin{equation} \label{eq:sensitivity_lim_2}
     L_{em} \geq \frac {1.22^2 \alpha \pi h c  AMD \lambda_{l} d_{E}^2}{4 \epsilon \Delta t N_{obs} A_{APF} d_T^2}
\end{equation}

For the APF telescope, $A_{APF} =  \pi(1.2 m)^2 = 4.53 m^2$, and $\epsilon = 5\%$. Exposure times range from several seconds to 20 minutes, with a median exposure time of $t = 730s$. Taking the median values of distance of the star in our target sample, 78.47 ly, and corresponding value for AMD of 414.78, and wavelength $\lambda_{l}= 5000$\,\AA, we find that our laser search algorithm would be able to detect lasers of luminosity \LemkW\ kW or greater. This derivation assumes a laser is invariant in time. We are also sensitive to pulsed lasers if the integrated laser flux in each observation of the host star reaches the detection threshold.

\subsection{Sensitivity Advantage of Residual Search}
\label{subsec:sensitivityadvantage}

The sensitivity of our search is increased when searching through the residuals rather than the spectra themselves. This is due to the fact that a search of the residuals catches emission lines which fall into absorption features in the target spectrum, and is evidenced by the greater number of flagged events arising from the residual search. The importance of this effect varies between stellar types, with the strongest effect in stars with the most absorption lines. A  search conducted on the residual of a very cool, very hot, or rapidly rotating star with a smooth featureless spectrum will be very similar to a search conducted on the spectrum itself. Conversely, for main sequence stars with deep features the residual search is more sensitive.  

We assess the improved sensitivity of searching the residuals by considering a representative emission feature that happens to fall within an absorption feature of depth $d$ below the continuum. In order to detect an event in the spectrum itself, we require the minimum height of the event to be greater by a value $d$  than if it did not fall into the absorption feature. Equation \ref{eq:requirement} becomes

\begin{equation} \label{eq:requirement_in_abs_featu}
    N_{det} \geq \alpha (AMD) + d
\end{equation}

Carrying this additional term through into Equation \ref{eq:sensitivity_lim_2}, we find the effect is an additional additive term to the luminosity required for detection. If we define $\beta$ such that Equation \ref{eq:sensitivity_lim_2} becomes $L_{em} \geq \alpha AMD / \beta$, the luminosity requirement within an absorption feature becomes: 

\begin{equation} \label{eq:sensetivity_abs_feature}
     L_{em} \geq \frac {\alpha (AMD) + d}{\beta} 
\end{equation}

When an emission line falls into an absorption feature, we effectively require the laser luminosity be greater by an additive factor of $d/\beta$. In regions of a spectrum away from stellar absorption features, searching the residuals provides no additional sensitivity. We define regions affected by absorption features (regions in which removing the stellar background meaningfully shifts the pixel values toward the continuum) as pixels whose values are greater than three absolute median deviations below the continuum. Pixels closer than this to the continuum are still shifted during the creation of the residual, but by an amount that is within the stellar noise. Using the residuals meaningfully increases the effective sensitivity only for emission lines which fall on pixels lower than this threshold. For the median spectrum, 53.9\% of pixels meet this requirement, and of those pixels the median depth is 0.12 below the continuum. Using our median $\beta$ value of $0.00278$, this corresponds to a median effective increase in sensitivity of 1.2\% over all pixels which fall along an absorption feature. The effect is most important near the centers of deep features; pixels in the center of absorption features reach a maximum of 0.96 below the continuum, corresponding to an effective increase of 9.7\%.

\section{Discussion and Conclusions} \label{sec:conclusion}

As part of the BL search for technosignatures, we have developed a pipeline for processing stellar spectra to produce residuals and searching the residuals for laser lines. We have adapted the \texttt{SpecMatch-Emp} model to operate on APF spectra in order to subtract away stellar flux from each target spectrum, allowing for a modestly more sensitive search near stellar absorption features. We acknowledge that the improvement in sensitivity is small for most potential lasers, and thus conclude that the most important way we increase our sensitivity in this work is through our consecutive observation strategy. We have presented the results of applying this algorithm to a target set of \NSearched\ stellar targets, using \NSpectraSearched\ APF spectra in the region 4997.10 -- 5899.99\,\AA. An injection and recovery test of Gaussian signals verifies that we can accurately recover their wavelengths and intensities. We find that six candidates pass our vetting tests when the spectra are searched directly, and 49 additional candidates (55 total) pass our tests when the residuals are searched. We vet these candidates manually, and have visually classified 11 as night sky emission lines, 9 as false positives from high rotational velocity stars, 21 as false positives from low metallicity stars, and 14 as artifacts of poor \texttt{SpecMatch-Emp} absorption feature subtraction. This leaves no candidates of likely non-astrophysical origin. 

From our non-detection, we impose an upper limit on the occurrence rate of persistent lasers directed towards Earth from bright nearby stars. Using Poisson statistics, the upper limit on laser emission lines at 95\%\ confidence with zero detected is 2.99. We thus conclude that less than 2.99 per \NSearched\ stars, or less than 0.78\%, of stars similar to those in our target set host persistent lasers above our detection threshold directed towards Earth. We are not sensitive to time-dependent lasers, and so can make no conclusion about temporary laser emission. 

We also can inherently only detect emission directed towards Earth. If we assume an emitting telescope with diameter similar to the APF (2.4m), the laser will fill only about 1 in $10^{15}$ of the sky area as viewed from the point of emission due to the geometric spreading of the laser. We note that this does not mean the pointing must be precise in order to fall on the APF if directed at Earth, but rather that this is a very small fraction of all three-dimensional space. Our example laser would spread to an area the size of the Earth by about 0.005\,ly, and an area the size of Earth's orbit by about 125\,ly.  Thus, we are much more sensitive to lasers directed intentionally towards Earth and to very high intensity lasers emitted from small apertures (and thus with wide spreading angles). We are less sensitive to lasers that happen to be inadvertently directed through our location. As mentioned in section \ref{subsec:targets}, we include a set of known transiting targets in part because intentional communication may be more likely along the ecliptic plane. Because of the lack of any model of the occurrence rate, technological advancement, and motives of any intelligent life beyond Earth, we do not make a claim on the implication of our non-detection for the occurrence rate of extraterrestrial intelligence, and only on the occurrence of detectable lasers.

Our laser search algorithm is designed to be agnostic to the source of any lasers we might detect, but we note here that our algorithm is more sensitive to certain potential sources than to others. A previously unknown astrophysical source of stellar laser emission would be well suited to detection via our method, as would emission originating from planets or spacecraft that are moving slowly relative to their host star. Our search method is less sensitive to lasers originating from planets or spacecraft which orbit their host stars quickly, because the apparent wavelength of the laser could change on timescales similar to the intervals between exposures of the same star. Our spectra are analyzed in the stellar rest frame, and so in checking that events persist across observations we must assume the laser emitted has the same velocity relative to its host star in each exposure. Due to the short ($\sim 20$\,min) interval between exposures, the typical wavelength shift between exposures would generally be absorbed in rounding each peak location to the nearest pixel value, but for short period planets the change in position could be up to several pixels. If an intelligent society were to attempt direct communication via laser emission, it is possible that this civilization would also correct for changes in the relative velocity of the host planet and host star in order to be more easily detected. 

We have determined that a major source of systematic uncertainty in our analysis comes from the production of the residuals when a target star's properties are near or outside the bounds of the \texttt{SpecMatch-Emp} library. Thus a priority for future work is to expand the library through inclusion of stars with a wider range of properties, in particular stars with large radii and very cold or hot stars. 

We have shown that once an event is observed in the residual spectrum, we can accurately recover its initial wavelength and intensity. We would be able to detect a \LemkW\,kW laser at the median distance of a star in our dataset. As a benchmark, the most powerful lasers on Earth can reach, for extremely brief durations of trillionths of a second, luminosities of several Petawatts \citep{Obayashi2015}. Common laser guide stars used in adaptive optics reach luminosities of tens of watts.

\subsection{Comparison to Previous Studies}

\subsubsection{Lipman et~al.~(2019)}
\cite{Lipman2019} searched 177 APF observations of Boyajian’s star for laser emission, using a similar set of criteria to classify the intensity and shape of events as is used in this work. They also used spectra from the APF, analyzing the full APF wavelength region $3740 - 9700$\,\AA\ with a median SNR of 30 per pixel, compared to our more limited spectral range of the \texttt{SpecMatch-Emp} library. Boyajian's star is a rapidly-rotating F-type star with no narrow spectral lines. Stars of this type benefit less from the search on residuals compared to slowly-rotating FGKM stars. Our work is broader in searching the spectra of 388 stars across a wide range of spectral types, rather than only one star. \citeauthor{Lipman2019} conducted their search pixel-by-pixel, imposing two conditions: that the width of the event is close to the PSF of the telescope, and that the wavelength does not match known atmospheric or stellar emission lines. They also visually inspected the 2D echelle spectra to reject cosmic ray hits. The authors conclude that their search has a sensitivity limit equivalent to detecting a 7.37\,MW  laser near Boyajian’s Star (1470\,ly away) \footnote{\cite{Lipman2019} actually report a detection threshold of 24\,MW at the distance of Boyajian's Star, but this appears to be due to a mistake converting nm to \AA\ during the calculation. Correcting for this gives 7.37\,MW.}. Using Equation ~\ref{eq:sensitivity_lim_2} above for the same distance, our algorithm would detect a laser with luminosity 4.77\,MW. We report a modestly higher sensitivity than that achieved by \citeauthor{Lipman2019}, but of the same order of magnitude. \cite{Lipman2019} do not report an upper bound on emitter prevalence based on their non-detection.

\subsubsection{Tellis et~al.~(2015)}
\cite{TellisMarcy2015} searched 2796 Keck HIRES spectra for laser emission, including a subset of the California Planet Search stars and 1368 KOIs with V $<$ 14.2. Their target set consisted primarily of FGKM main-sequence stars and some subgiants. Though our target set is significantly smaller in number, it is unique in diversity, comprising stars outside of the typical planet search stellar property range and allowing us to investigate a previously unexplored region of parameter space. \citeauthor{TellisMarcy2015}'s spectra were taken with varying exposure times to maintain an SNR of 100-200 independent of stellar magnitude.
They searched the wavelength region between 3640 and 7890\,\AA, about four times the region searched in this work. Their search was conducted pixel-by-pixel in the raw (2D) spectra and consisted of tests on the intensity (SNR) of the potential laser emission compared to surrounding pixels, and on the goodness-of-fit to the telescope PSF. They rejected atmospheric lines using a catalog of known night sky lines. They visually inspected 15604 events which met their initial criteria, and rejected all as possible laser emission. Their work searched the region on the 2D echellogram that is absent of light from the target star, a range of $2 - 7$\arcsec\ from the target star. Because \cite{TellisMarcy2015} omitted the region coincident with the star and  did not require potential laser emission to compete with stellar background flux, they were able to achieve high sensitivities in the region they did search. However, this came at the expense of missing possible emission from planets near a host star. They report that their algorithm could detect a 90\,W or more powerful laser within 100\,ly of Earth, or a 1\,kW or more powerful laser within 1000\,ly, if the laser is $60-200$\,AU, or in the second case $2000-7000$\,AU, from nearby Kepler stars. Using Equation ~\ref{eq:sensitivity_lim_2} above for a distance of 100\,ly, our algorithm would detect a laser with luminosity of 130 kW. \cite{TellisMarcy2015} achieve their higher sensitivity by restricting themselves to the spatial region well away from contaminating stellar light. \cite{TellisMarcy2015} also do not report an upper bound on emitter prevalence based on their nondetection.

\subsubsection{Tellis et~al.~(2017)}

\cite{TellisMarcy2017} searched the reduced spectra of 5600 Keck-HIRES stars in the same $3640 - 7890$\,\AA range as their 2015 work, with a spectral resolution of 60000. They visually inspected 5023 events which met their initial criteria, and rejected all as possible laser emission. Their targets were mostly main-sequence stars and subgiants of FGKM spectral type, including many California Planet Search stars (67708 total spectra), and many transiting exoplanet search candidates from COROT, Kepler and the HATNET program. They did not exclude spectra from galaxies, nebulae, etc. Similarly to their 2015 survey, their 2017 target set is larger and complementary to the stars searched in our work. They conducted their search by imposing the following criteria on events: that the width matched the instrumental profile and was greater than the telescope PSF, and that the intensity above the continuum was greater than the effective noise. They rejected candidates that matched known Balmer series stellar emission lines and common night sky emission lines, as well as grazing cosmic ray hits which passed the earlier tests. This search was hindered by the numerous spectral absorption features and iodine lines. The authors describe stellar absorption features as their predominant source of noise; by conducting our search on the residual spectra we overcome this issue. \cite{TellisMarcy2017} note a difficulty in searching spectra with signal-to-noise less than 10 per pixel. Our search shares a similar difficulty when searching low SNR spectra. 

Because they compensated for target distance by increasing the exposure time to maintain similar SNR values (always $>$ 25, and generally over 100) across targets, their sensitivity limit is not dependent on target distance. However, they report a sensitivity limit which depends strongly on laser wavelength and host spectral type, ranging from 3\,kW to 13\,MW (and on the order of 1\,MW for most spectral types). They report the lowest sensitivities for later spectral types with many molecular absorption features which obscure individual emission lines and make the continuum less well defined. The wavelength dependence of their sensitivity comes from the fact that some spectral types have a smoother spectrum at higher wavelengths, partially mitigating the effects of local absorption features in those regions. For some spectral types the continuum flux value can vary across the spectrum, requiring higher intensity lasers for detection in higher continuum regions. If we take as an example a laser at the median distance of a star in our target set, 78.47\,ly, our sensitivity reported in Section \ref{subsec:sensetivity} falls  within \cite{TellisMarcy2017}'s reported range. Despite relying on a different set of assumptions about the nature of technologically advanced life, \cite{TellisMarcy2017} report a similar upper limit on emitter prevalence to that approximated in Section~\ref{sec:conclusion}, calculating a value of $0.1\%$.

\subsection{Future Directions} \label{future}

As the number of routines designed to search optical spectra grows, one possible next step is application of existing codes to archived high-resolution spectra. Precise radial velocity surveys that use slit-fed high-resolution spectrographs share a similar format with APF data and are amenable to a similar type of laser search. The next generation of high-resolution spectrographs are fiber fed, and the resulting spectra can be searched for laser lines with a similar strategy,but the methods for rejecting candidates will differ. Many of these spectrographs, such as the Keck Planet Finder, Habitable Planet Finder, NEID, EXPRES and ESPRESSO, use an image slicer resulting in several traces per order in the 2D echellogram. Several unique measures of the same wavelength are produced. The slices are extracted separately such that emission lines originating from a star's proximity occur in all of the order traces at that wavelength. Cosmic ray hits will be rejected more easily. The sky fiber, placed tens of arcseconds away from the target star, offers a powerful tool for rejecting night sky emission lines by collecting an independent measure of the night sky.

Another path forward will be to establish searches on archived high-resolution spectra from large surveys such as APOGEE and Gaia DR3. The focus of these searches will be expanding the sample size of stars searched from thousands to millions of stars. Care will have to be given to understanding the automated processing pipelines of these surveys, but the sheer number of stars and spectra available to search for technosignatures will justify the effort.

\bigskip
Acknowledgments: We thank The Breakthrough Listen Initiative, and everyone who has provided support for this project. 
We thank Gloria and Ken Levy for support of the Automated Planet Finder Spectrometer. We are grateful to the Berkeley SETI Research Center (BSRC) for their support of this research. We thank the staff and students of Breakthrough Listen for their support of SETI data collection and making this data publicly available at \url{https://seti.berkeley.edu/opendata}. Breakthrough Listen is managed by the Breakthrough Initiatives, sponsored by the Breakthrough Prize Foundation. Research at Lick Observatory is partially supported by a generous gift from Google.

AZ and ZK were funded as participants in the Berkeley SETI Research Center Research Experience for Undergraduates Site, supported by the National Science Foundation under Grant No.~1950897.

This work has made use of data from the European Space Agency (ESA) mission
{\it Gaia} (\url{https://www.cosmos.esa.int/gaia}), processed by the {\it Gaia}
Data Processing and Analysis Consortium (DPAC,
\url{https://www.cosmos.esa.int/web/gaia/dpac/consortium}). Funding for the DPAC
has been provided by national institutions, in particular the institutions
participating in the {\it Gaia} Multilateral Agreement.

We thank the referee for helpful comments which improved the final version of this paper.


%

\vspace{5mm}
\facilities{Automated Planet Finder (Levy), Lick Observatory}


\software{SpecMatch-Emp (\url{https://github.com/samuelyeewl/specmatch-emp/}),  Isoclassify 
(\url{https://github.com/danxhuber/isoclassify})}



\appendix

\section{Sample results Table (\texttt{SpecMatch-Emp})}

\begin{table}[H]
\centering
\begin{tabular}{ccccccc}
$Name$    & $T_{eff} [K]$  & $R [R_\odot]$    & $[Fe/H] [dex]$    & $M [M_\odot]$   & $ log_{10}(age) $ & $log(g)$        \\ \hline
HIP105341 & $4129 \pm 72$  & $0.65 \pm 0.070$ & $0.05 \pm 0.090$  & $0.68 \pm 0.60$ & $9.76 \pm 0.23$   & $4.66 \pm 0.09$ \\
HIP105668 & $6648 \pm 126$ & $1.63 \pm 0.340$ & $-0.13 \pm 0.090$ & $1.40 \pm 0.60$ & $9.37 \pm 0.23$   & $4.05 \pm 0.09$ \\
HIP105769 & $6624 \pm 126$ & $2.03 \pm 0.340$ & $-0.09 \pm 0.090$ & $1.46 \pm 0.60$ & $9.34 \pm 0.23$   & $4.00 \pm 0.09$ \\
HIP105860 & $6487 \pm 126$ & $2.48 \pm 0.340$ & $0.05 \pm 0.090$  & $1.62 \pm 0.60$ & $9.28 \pm 0.23$   & $3.86 \pm 0.09$ \\
HIP105932 & $3675 \pm 72$  & $0.43 \pm 0.070$ & $-0.24 \pm 0.090$ & $0.45 \pm 0.60$ & $9.63 \pm 0.23$   & $4.83 \pm 0.09$ \\
HIP106400 & $4385 \pm 72$  & $0.68 \pm 0.070$ & $-0.09 \pm 0.090$ & $0.68 \pm 0.60$ & $9.60 \pm 0.23$   & $4.65 \pm 0.09$ \\
HIP106481 & $5006 \pm 126$ & $4.43 \pm 0.340$ & $-0.02 \pm 0.090$ & $1.13 \pm 0.60$ & $9.84 \pm 0.23$   & $3.22 \pm 0.09$ \\
HIP10670  & $6702 \pm 126$ & $1.52 \pm 0.340$ & $-0.19 \pm 0.090$ & $1.34 \pm 0.60$ & $9.39 \pm 0.23$   & $4.10 \pm 0.09$ \\
HIP106897 & $6646 \pm 126$ & $1.63 \pm 0.340$ & $-0.13 \pm 0.090$ & $1.40 \pm 0.60$ & $9.37 \pm 0.23$   & $4.04 \pm 0.09$ \\
HIP107346 & $4127 \pm 72$  & $0.65 \pm 0.070$ & $-0.04 \pm 0.090$ & $0.66 \pm 0.60$ & $9.81 \pm 0.23$   & $4.66 \pm 0.09$ \\
HIP107350 & $5954 \pm 126$ & $1.05 \pm 0.340$ & $-0.02 \pm 0.090$ & $1.06 \pm 0.60$ & $9.38 \pm 0.23$   & $4.45 \pm 0.09$ \\
HIP107788 & $6669 \pm 126$ & $1.57 \pm 0.340$ & $-0.16 \pm 0.090$ & $1.37 \pm 0.60$ & $9.38 \pm 0.23$   & $4.08 \pm 0.09$ \\
HIP107975 & $6248 \pm 126$ & $1.48 \pm 0.340$ & $-0.41 \pm 0.090$ & $1.06 \pm 0.60$ & $9.74 \pm 0.23$   & $4.04 \pm 0.09$ \\
HIP108028 & $4969 \pm 126$ & $0.76 \pm 0.070$ & $-0.04 \pm 0.090$ & $0.79 \pm 0.60$ & $9.85 \pm 0.23$   & $4.52 \pm 0.09$ \\
HIP108036 & $6600 \pm 126$ & $2.00 \pm 0.340$ & $-0.09 \pm 0.090$ & $1.45 \pm 0.60$ & $9.35 \pm 0.23$   & $4.01 \pm 0.09$ \\
HIP108092 & $3922 \pm 72$  & $0.66 \pm 0.070$ & $-0.07 \pm 0.090$ & $0.60 \pm 0.60$ & $9.65 \pm 0.23$   & $4.69 \pm 0.09$ \\
HIP108156 & $4983 \pm 126$ & $0.79 \pm 0.070$ & $-0.00 \pm 0.090$ & $0.79 \pm 0.60$ & $10.00 \pm 0.23$  & $4.47 \pm 0.09$ \\
HIP108506 & $4892 \pm 126$ & $4.28 \pm 0.340$ & $0.12 \pm 0.090$  & $1.18 \pm 0.60$ & $9.81 \pm 0.23$   & $3.32 \pm 0.09$ \\
HIP1086   & $6430 \pm 126$ & $1.58 \pm 0.340$ & $-0.14 \pm 0.090$ & $1.30 \pm 0.60$ & $9.40 \pm 0.23$   & $4.11 \pm 0.09$ \\
HIP108782 & $3783 \pm 72$  & $0.58 \pm 0.070$ & $0.02 \pm 0.090$  & $0.60 \pm 0.60$ & $9.87 \pm 0.23$   & $4.69 \pm 0.09$ \\
HIP109378 & $5357 \pm 126$ & $1.51 \pm 0.340$ & $0.15 \pm 0.090$  & $1.01 \pm 0.60$ & $9.98 \pm 0.23$   & $4.05 \pm 0.09$ \\
HIP109388 & $3471 \pm 72$  & $0.44 \pm 0.070$ & $0.10 \pm 0.090$  & $0.44 \pm 0.60$ & $9.64 \pm 0.23$   & $4.81 \pm 0.09$ \\
HIP109427 & $6688 \pm 126$ & $1.54 \pm 0.340$ & $-0.18 \pm 0.090$ & $1.35 \pm 0.60$ & $9.39 \pm 0.23$   & $4.09 \pm 0.09$ \\
HIP109474 & $6371 \pm 126$ & $1.48 \pm 0.340$ & $-0.12 \pm 0.090$ & $1.26 \pm 0.60$ & $9.39 \pm 0.23$   & $4.17 \pm 0.09$ \\
HIP109555 & $3586 \pm 72$  & $0.48 \pm 0.070$ & $0.06 \pm 0.090$  & $0.49 \pm 0.60$ & $9.71 \pm 0.23$   & $4.76 \pm 0.09$ \\
HIP109638 & $3480 \pm 72$  & $0.40 \pm 0.070$ & $-0.01 \pm 0.090$ & $0.40 \pm 0.60$ & $9.68 \pm 0.23$   & $4.84 \pm 0.09$ \\
HIP109822 & $4960 \pm 126$ & $3.03 \pm 0.340$ & $-0.10 \pm 0.090$ & $0.97 \pm 0.60$ & $9.99 \pm 0.23$   & $3.54 \pm 0.09$ \\
HIP109857 & $6585 \pm 126$ & $2.13 \pm 0.340$ & $-0.06 \pm 0.090$ & $1.49 \pm 0.60$ & $9.33 \pm 0.23$   & $3.97 \pm 0.09$ \\
HIP109926 & $5342 \pm 126$ & $0.82 \pm 0.070$ & $0.03 \pm 0.090$  & $0.89 \pm 0.60$ & $9.51 \pm 0.23$   & $4.54 \pm 0.09$ \\
HIP4845   & $3976 \pm 72$  & $0.62 \pm 0.070$ & $-0.16 \pm 0.090$ & $0.61 \pm 0.60$ & $9.83 \pm 0.23$   & $4.68 \pm 0.09$ \\
HIP4849   & $4700 \pm 126$ & $0.75 \pm 0.070$ & $-0.08 \pm 0.090$ & $0.76 \pm 0.60$ & $9.96 \pm 0.23$   & $4.55 \pm 0.09$ \\
HIP4856   & $3430 \pm 72$  & $0.41 \pm 0.070$ & $-0.01 \pm 0.090$ & $0.42 \pm 0.60$ & $9.64 \pm 0.23$   & $4.84 \pm 0.09$ \\
HIP4872   & $3654 \pm 72$  & $0.54 \pm 0.070$ & $0.08 \pm 0.090$  & $0.54 \pm 0.60$ & $9.80 \pm 0.23$   & $4.72 \pm 0.09$ \\
HIP4907   & $5328 \pm 126$ & $0.93 \pm 0.070$ & $-0.07 \pm 0.090$ & $0.86 \pm 0.60$ & $9.89 \pm 0.23$   & $4.45 \pm 0.09$ \\
HIP49081  & $5682 \pm 126$ & $1.24 \pm 0.340$ & $0.13 \pm 0.090$  & $1.02 \pm 0.60$ & $9.93 \pm 0.23$   & $4.22 \pm 0.09$ \\
HIP49127  & $4882 \pm 126$ & $1.00 \pm 0.070$ & $0.03 \pm 0.090$  & $0.81 \pm 0.60$ & $9.89 \pm 0.23$   & $4.42 \pm 0.09$ \\
HIP50372  & $6682 \pm 126$ & $1.54 \pm 0.340$ & $-0.18 \pm 0.090$ & $1.35 \pm 0.60$ & $9.39 \pm 0.23$   & $4.09 \pm 0.09$ \\
HIP50384  & $6031 \pm 126$ & $1.33 \pm 0.340$ & $-0.32 \pm 0.090$ & $1.00 \pm 0.60$ & $9.71 \pm 0.23$   & $4.18 \pm 0.09$ \\
HIP50485  & $5006 \pm 126$ & $3.04 \pm 0.340$ & $0.15 \pm 0.090$  & $1.12 \pm 0.60$ & $9.90 \pm 0.23$   & $3.59 \pm 0.09$ \\
HIP51384  & $6646 \pm 126$ & $1.62 \pm 0.340$ & $-0.13 \pm 0.090$ & $1.40 \pm 0.60$ & $9.37 \pm 0.23$   & $4.05 \pm 0.09$ \\
HIP51459  & $5953 \pm 126$ & $1.15 \pm 0.340$ & $-0.07 \pm 0.090$ & $1.05 \pm 0.60$ & $9.64 \pm 0.23$   & $4.32 \pm 0.09$ \\
HIP51502  & $6623 \pm 126$ & $1.63 \pm 0.340$ & $-0.12 \pm 0.090$ & $1.41 \pm 0.60$ & $9.37 \pm 0.23$   & $4.04 \pm 0.09$ \\
HIP51819  & $5262 \pm 126$ & $0.89 \pm 0.070$ & $0.05 \pm 0.090$  & $0.87 \pm 0.60$ & $9.62 \pm 0.23$   & $4.50 \pm 0.09$ \\
HIP51933  & $6033 \pm 126$ & $1.24 \pm 0.340$ & $-0.19 \pm 0.090$ & $1.04 \pm 0.60$ & $9.70 \pm 0.23$   & $4.25 \pm 0.09$ \\
HIP52316  & $5009 \pm 126$ & $3.54 \pm 0.340$ & $0.04 \pm 0.090$  & $1.18 \pm 0.60$ & $9.80 \pm 0.23$   & $3.41 \pm 0.09$
\end{tabular}
\caption{Selected columns and rows from the results of \texttt{SpecMatch-Emp} before isochrone analysis. The full file includes 902 rows and 58 columns, including the stellar property results and uncertainties for each star, as well as various diagnostic information such as flags for spectroscopic binaries, the chi-squared value of the fit to the \texttt{SpecMatch-Emp} matched spectra, and any warning messages.}
\label{tab:SM_results_sample}
\end{table}

\section{Sample Results Table (\texttt{SpecMatch-Emp} and Isochhrone Analysis)}
\label{subsec:app-isochrone}

\begin{table}[H]
\centering
\begin{tabular}{ccccccc}
$Name$    & $T_{eff} [K]$               & $R [R_\odot]$                 & $[Fe/H] [dex]$                 & $M [M_\odot]$                 & $log_{10}(age) $               & $log(g)$                      \\ \hline
HIP105341 & $4177^{ + 42 } _ { - 48}$   & $0.64^{ + 0.02 } _ { - 0.02}$ & $0.06^{ + 0.08 } _ { - 0.08}$  & $0.63^{ + 0.02 } _ { - 0.02}$ & $9.72^{ + 6.58 } _ { - 6.44}$  & $4.63^{ + 0.03 } _ { - 0.02}$ \\
HIP105668 & $6667^{ + 117 } _ { - 110}$ & $2.31^{ + 0.06 } _ { - 0.06}$ & $-0.10^{ + 0.07 } _ { - 0.10}$ & $1.53^{ + 0.05 } _ { - 0.05}$ & $1.70^{ + 0.22 } _ { - 0.18}$  & $3.89^{ + 0.02 } _ { - 0.02}$ \\
HIP105769 & $6743^{ + 117 } _ { - 117}$ & $1.62^{ + 0.03 } _ { - 0.03}$ & $-0.07^{ + 0.08 } _ { - 0.10}$ & $1.36^{ + 0.04 } _ { - 0.04}$ & $1.76^{ + 0.29 } _ { - 0.29}$  & $4.15^{ + 0.02 } _ { - 0.02}$ \\
HIP105860 & $6460^{ + 113 } _ { - 133}$ & $1.61^{ + 0.04 } _ { - 0.04}$ & $0.03^{ + 0.08 } _ { - 0.08}$  & $1.33^{ + 0.04 } _ { - 0.03}$ & $2.19^{ + 0.38 } _ { - 0.37}$  & $4.15^{ + 0.02 } _ { - 0.02}$ \\
HIP105932 & $3784^{ + 43 } _ { - 40}$   & $0.43^{ + 0.01 } _ { - 0.01}$ & $-0.17^{ + 0.07 } _ { - 0.09}$ & $0.42^{ + 0.01 } _ { - 0.01}$ & $11.09^{ + 5.97 } _ { - 6.88}$ & $4.80^{ + 0.01 } _ { - 0.01}$ \\
HIP106400 & $4401^{ + 62 } _ { - 62}$   & $0.65^{ + 0.02 } _ { - 0.02}$ & $-0.08^{ + 0.08 } _ { - 0.08}$ & $0.66^{ + 0.02 } _ { - 0.02}$ & $9.33^{ + 6.65 } _ { - 6.20}$  & $4.63^{ + 0.02 } _ { - 0.02}$ \\
HIP106481 & $4972^{ + 97 } _ { - 76}$   & $6.37^{ + 1.44 } _ { - 0.74}$ & $-0.02^{ + 0.07 } _ { - 0.10}$ & $1.86^{ + 0.27 } _ { - 0.32}$ & $1.12^{ + 0.80 } _ { - 0.32}$  & $3.07^{ + 0.08 } _ { - 0.12}$ \\
HIP10670  & $6748^{ + 117 } _ { - 113}$ & $2.28^{ + 0.07 } _ { - 0.06}$ & $-0.14^{ + 0.07 } _ { - 0.10}$ & $1.52^{ + 0.05 } _ { - 0.05}$ & $1.68^{ + 0.22 } _ { - 0.17}$  & $3.90^{ + 0.03 } _ { - 0.02}$ \\
HIP106897 & $6683^{ + 125 } _ { - 113}$ & $1.68^{ + 0.04 } _ { - 0.04}$ & $-0.13^{ + 0.08 } _ { - 0.10}$ & $1.34^{ + 0.04 } _ { - 0.03}$ & $2.01^{ + 0.27 } _ { - 0.27}$  & $4.12^{ + 0.02 } _ { - 0.02}$ \\
HIP107346 & $4180^{ + 51 } _ { - 36}$   & $0.61^{ + 0.02 } _ { - 0.02}$ & $-0.04^{ + 0.07 } _ { - 0.07}$ & $0.62^{ + 0.02 } _ { - 0.02}$ & $9.71^{ + 6.59 } _ { - 6.43}$  & $4.65^{ + 0.02 } _ { - 0.03}$ \\
HIP107350 & $5978^{ + 85 } _ { - 85}$   & $0.99^{ + 0.02 } _ { - 0.02}$ & $-0.06^{ + 0.08 } _ { - 0.08}$ & $1.04^{ + 0.03 } _ { - 0.04}$ & $1.22^{ + 1.45 } _ { - 0.81}$  & $4.46^{ + 0.02 } _ { - 0.02}$ \\
HIP107788 & $6647^{ + 127 } _ { - 127}$ & $1.75^{ + 0.07 } _ { - 0.07}$ & $-0.16^{ + 0.10 } _ { - 0.09}$ & $1.34^{ + 0.04 } _ { - 0.04}$ & $2.13^{ + 0.28 } _ { - 0.27}$  & $4.08^{ + 0.03 } _ { - 0.03}$ \\
HIP107975 & $6273^{ + 117 } _ { - 130}$ & $1.65^{ + 0.04 } _ { - 0.04}$ & $-0.40^{ + 0.09 } _ { - 0.09}$ & $1.11^{ + 0.09 } _ { - 0.06}$ & $4.83^{ + 1.05 } _ { - 1.41}$  & $4.04^{ + 0.04 } _ { - 0.03}$ \\
HIP108028 & $5071^{ + 103 } _ { - 97}$  & $0.78^{ + 0.03 } _ { - 0.03}$ & $-0.05^{ + 0.09 } _ { - 0.09}$ & $0.80^{ + 0.04 } _ { - 0.04}$ & $6.31^{ + 6.08 } _ { - 4.24}$  & $4.56^{ + 0.03 } _ { - 0.04}$ \\
HIP108036 & $6571^{ + 121 } _ { - 123}$ & $1.70^{ + 0.05 } _ { - 0.04}$ & $-0.10^{ + 0.09 } _ { - 0.10}$ & $1.34^{ + 0.04 } _ { - 0.03}$ & $2.17^{ + 0.33 } _ { - 0.31}$  & $4.10^{ + 0.03 } _ { - 0.03}$ \\
HIP108092 & $4097^{ + 57 } _ { - 61}$   & $0.62^{ + 0.02 } _ { - 0.02}$ & $0.03^{ + 0.08 } _ { - 0.09}$  & $0.61^{ + 0.01 } _ { - 0.01}$ & $13.75^{ + 4.28 } _ { - 6.82}$ & $4.64^{ + 0.02 } _ { - 0.02}$ \\
HIP108156 & $5085^{ + 88 } _ { - 79}$   & $0.80^{ + 0.04 } _ { - 0.03}$ & $-0.02^{ + 0.09 } _ { - 0.09}$ & $0.81^{ + 0.03 } _ { - 0.04}$ & $7.37^{ + 5.98 } _ { - 4.74}$  & $4.54^{ + 0.04 } _ { - 0.04}$ \\
HIP108506 & $4786^{ + 78 } _ { - 65}$   & $3.15^{ + 0.08 } _ { - 0.08}$ & $0.08^{ + 0.10 } _ { - 0.10}$  & $0.98^{ + 0.14 } _ { - 0.09}$ & $11.32^{ + 4.84 } _ { - 4.36}$ & $3.43^{ + 0.06 } _ { - 0.04}$ \\
HIP1086   & $6574^{ + 112 } _ { - 111}$ & $1.68^{ + 0.03 } _ { - 0.03}$ & $-0.12^{ + 0.11 } _ { - 0.09}$ & $1.33^{ + 0.03 } _ { - 0.03}$ & $2.21^{ + 0.30 } _ { - 0.28}$  & $4.11^{ + 0.02 } _ { - 0.02}$ \\
HIP108782 & $3881^{ + 49 } _ { - 46}$   & $0.56^{ + 0.01 } _ { - 0.01}$ & $0.12^{ + 0.07 } _ { - 0.07}$  & $0.55^{ + 0.01 } _ { - 0.01}$ & $12.48^{ + 5.13 } _ { - 6.99}$ & $4.68^{ + 0.01 } _ { - 0.01}$ \\
HIP109378 & $5421^{ + 110 } _ { - 101}$ & $1.12^{ + 0.02 } _ { - 0.02}$ & $0.14^{ + 0.10 } _ { - 0.07}$  & $0.92^{ + 0.05 } _ { - 0.04}$ & $13.04^{ + 3.19 } _ { - 3.10}$ & $4.30^{ + 0.03 } _ { - 0.03}$ \\
HIP109388 & $3615^{ + 42 } _ { - 43}$   & $0.46^{ + 0.01 } _ { - 0.01}$ & $0.21^{ + 0.07 } _ { - 0.07}$  & $0.45^{ + 0.01 } _ { - 0.01}$ & $11.98^{ + 5.44 } _ { - 6.90}$ & $4.76^{ + 0.01 } _ { - 0.01}$ \\
HIP109427 & $6703^{ + 120 } _ { - 119}$ & $2.05^{ + 0.19 } _ { - 0.18}$ & $-0.16^{ + 0.07 } _ { - 0.09}$ & $1.44^{ + 0.08 } _ { - 0.07}$ & $1.88^{ + 0.32 } _ { - 0.25}$  & $3.96^{ + 0.06 } _ { - 0.06}$ \\
HIP109474 & $6615^{ + 78 } _ { - 114}$  & $1.67^{ + 0.03 } _ { - 0.03}$ & $-0.07^{ + 0.09 } _ { - 0.09}$ & $1.34^{ + 0.04 } _ { - 0.03}$ & $2.08^{ + 0.29 } _ { - 0.28}$  & $4.12^{ + 0.02 } _ { - 0.02}$ \\
HIP109555 & $3742^{ + 42 } _ { - 44}$   & $0.52^{ + 0.01 } _ { - 0.01}$ & $0.19^{ + 0.07 } _ { - 0.06}$  & $0.51^{ + 0.01 } _ { - 0.01}$ & $12.76^{ + 4.96 } _ { - 6.95}$ & $4.71^{ + 0.01 } _ { - 0.01}$ \\
HIP109638 & $3540^{ + 61 } _ { - 52}$   & $0.37^{ + 0.01 } _ { - 0.01}$ & $0.08^{ + 0.06 } _ { - 0.09}$  & $0.36^{ + 0.01 } _ { - 0.01}$ & $10.08^{ + 6.46 } _ { - 6.63}$ & $4.84^{ + 0.01 } _ { - 0.01}$ \\
HIP109822 & $4970^{ + 91 } _ { - 86}$   & $2.78^{ + 0.07 } _ { - 0.07}$ & $-0.09^{ + 0.08 } _ { - 0.10}$ & $1.05^{ + 0.12 } _ { - 0.12}$ & $7.47^{ + 4.30 } _ { - 2.52}$  & $3.57^{ + 0.05 } _ { - 0.06}$ \\
HIP109857 & $6572^{ + 121 } _ { - 117}$ & $2.12^{ + 0.21 } _ { - 0.19}$ & $-0.05^{ + 0.07 } _ { - 0.10}$ & $1.47^{ + 0.09 } _ { - 0.08}$ & $1.88^{ + 0.36 } _ { - 0.28}$  & $3.94^{ + 0.07 } _ { - 0.06}$ \\
HIP109926 & $5293^{ + 90 } _ { - 87}$   & $0.80^{ + 0.01 } _ { - 0.01}$ & $-0.03^{ + 0.09 } _ { - 0.07}$ & $0.87^{ + 0.02 } _ { - 0.03}$ & $2.26^{ + 2.86 } _ { - 1.56}$  & $4.57^{ + 0.02 } _ { - 0.02}$ \\
HIP4845   & $4115^{ + 40 } _ { - 60}$   & $0.60^{ + 0.02 } _ { - 0.02}$ & $-0.06^{ + 0.08 } _ { - 0.08}$ & $0.59^{ + 0.01 } _ { - 0.01}$ & $13.60^{ + 4.38 } _ { - 6.90}$ & $4.66^{ + 0.02 } _ { - 0.02}$ \\
HIP4849   & $4771^{ + 97 } _ { - 96}$   & $0.68^{ + 0.02 } _ { - 0.02}$ & $-0.17^{ + 0.08 } _ { - 0.08}$ & $0.72^{ + 0.03 } _ { - 0.03}$ & $3.68^{ + 5.67 } _ { - 2.72}$  & $4.63^{ + 0.02 } _ { - 0.03}$ \\
HIP4856   & $3628^{ + 40 } _ { - 38}$   & $0.46^{ + 0.01 } _ { - 0.01}$ & $0.16^{ + 0.07 } _ { - 0.06}$  & $0.44^{ + 0.01 } _ { - 0.01}$ & $12.65^{ + 5.01 } _ { - 6.81}$ & $4.77^{ + 0.01 } _ { - 0.01}$ \\
HIP4872   & $3771^{ + 43 } _ { - 43}$   & $0.53^{ + 0.01 } _ { - 0.01}$ & $0.17^{ + 0.07 } _ { - 0.07}$  & $0.52^{ + 0.01 } _ { - 0.01}$ & $12.23^{ + 5.28 } _ { - 6.97}$ & $4.71^{ + 0.01 } _ { - 0.01}$ \\
HIP4907   & $5382^{ + 95 } _ { - 91}$   & $0.83^{ + 0.01 } _ { - 0.01}$ & $-0.10^{ + 0.08 } _ { - 0.09}$ & $0.86^{ + 0.03 } _ { - 0.04}$ & $4.54^{ + 3.90 } _ { - 2.81}$  & $4.54^{ + 0.02 } _ { - 0.03}$ \\
HIP49081  & $5662^{ + 117 } _ { - 114}$ & $1.20^{ + 0.04 } _ { - 0.04}$ & $0.11^{ + 0.10 } _ { - 0.08}$  & $1.00^{ + 0.05 } _ { - 0.05}$ & $8.53^{ + 2.81 } _ { - 2.50}$  & $4.28^{ + 0.04 } _ { - 0.04}$ \\
HIP49127  & $4948^{ + 101 } _ { - 97}$  & $0.78^{ + 0.03 } _ { - 0.03}$ & $0.01^{ + 0.09 } _ { - 0.09}$  & $0.79^{ + 0.04 } _ { - 0.04}$ & $8.64^{ + 6.31 } _ { - 5.50}$  & $4.55^{ + 0.04 } _ { - 0.04}$ \\
HIP50372  & $6702^{ + 121 } _ { - 117}$ & $2.08^{ + 0.21 } _ { - 0.19}$ & $-0.15^{ + 0.07 } _ { - 0.10}$ & $1.45^{ + 0.09 } _ { - 0.08}$ & $1.85^{ + 0.33 } _ { - 0.26}$  & $3.95^{ + 0.07 } _ { - 0.06}$ \\
HIP50384  & $6019^{ + 118 } _ { - 117}$ & $1.27^{ + 0.03 } _ { - 0.03}$ & $-0.32^{ + 0.07 } _ { - 0.09}$ & $0.97^{ + 0.06 } _ { - 0.05}$ & $7.74^{ + 2.10 } _ { - 1.94}$  & $4.21^{ + 0.04 } _ { - 0.04}$ \\
HIP50485  & $5053^{ + 107 } _ { - 90}$  & $2.40^{ + 0.05 } _ { - 0.04}$ & $0.13^{ + 0.10 } _ { - 0.11}$  & $1.18^{ + 0.06 } _ { - 0.08}$ & $5.69^{ + 1.48 } _ { - 0.92}$  & $3.75^{ + 0.03 } _ { - 0.03}$ \\
HIP51384  & $6823^{ + 123 } _ { - 105}$ & $1.90^{ + 0.05 } _ { - 0.05}$ & $-0.10^{ + 0.07 } _ { - 0.10}$ & $1.45^{ + 0.04 } _ { - 0.04}$ & $1.70^{ + 0.21 } _ { - 0.20}$  & $4.04^{ + 0.02 } _ { - 0.02}$ \\
HIP51459  & $5935^{ + 117 } _ { - 119}$ & $1.12^{ + 0.09 } _ { - 0.08}$ & $-0.08^{ + 0.08 } _ { - 0.10}$ & $1.01^{ + 0.06 } _ { - 0.06}$ & $5.11^{ + 2.45 } _ { - 2.22}$  & $4.34^{ + 0.06 } _ { - 0.06}$ \\
HIP51502  & $6561^{ + 130 } _ { - 127}$ & $1.39^{ + 0.04 } _ { - 0.04}$ & $-0.14^{ + 0.10 } _ { - 0.10}$ & $1.26^{ + 0.04 } _ { - 0.04}$ & $1.82^{ + 0.58 } _ { - 0.53}$  & $4.25^{ + 0.03 } _ { - 0.03}$ \\
HIP51819  & $5368^{ + 90 } _ { - 92}$   & $0.84^{ + 0.01 } _ { - 0.01}$ & $0.04^{ + 0.08 } _ { - 0.08}$  & $0.90^{ + 0.03 } _ { - 0.03}$ & $2.82^{ + 2.97 } _ { - 1.87}$  & $4.54^{ + 0.02 } _ { - 0.02}$ \\
HIP51933  & $6090^{ + 100 } _ { - 106}$ & $1.52^{ + 0.04 } _ { - 0.04}$ & $-0.16^{ + 0.08 } _ { - 0.08}$ & $1.13^{ + 0.06 } _ { - 0.07}$ & $4.89^{ + 1.75 } _ { - 1.17}$  & $4.13^{ + 0.03 } _ { - 0.04}$ \\
HIP52316  & $4913^{ + 90 } _ { - 80}$   & $2.79^{ + 0.06 } _ { - 0.05}$ & $-0.03^{ + 0.10 } _ { - 0.06}$ & $1.05^{ + 0.13 } _ { - 0.12}$ & $8.05^{ + 4.41 } _ { - 2.85}$  & $3.57^{ + 0.06 } _ { - 0.06}$ \\
\end{tabular}
\caption{Selected columns and rows from the results of \texttt{SpecMatch-Emp} and isochrone analysis. The full results file includes 902 rows and 58 columns, including the stellar property results and uncertainties for each star, as well as various diagnostic information such as flags for spectroscopic binaries, the chi-squared value of the fit to the \texttt{SpecMatch-Emp} matched spectra, and any warning messages.}
\label{tab:Iso_properties_sample}
\end{table}

\bibliography{works_cited}{}
\bibliographystyle{aasjournal}



\end{document}